\documentclass[twocolumn]{aastex631}
\usepackage{graphics,graphicx}
\hypersetup{urlcolor=blue}
\usepackage{natbib}
\usepackage[outdir=./]{epstopdf}
\usepackage{textcomp,gensymb}
\usepackage{xfrac}
\usepackage{xcolor}

\newcommand{\clustername}{MELANGE-3}
\newcommand{\association}{MELANGE-3}

\newcommand{\name}{Kepler-1928}
\newcommand{\starname}{Kepler-1928}
\newcommand{\planetname}{Kepler-1928\,b}

\newcommand{\starnametwo}{Kepler-970}
\newcommand{\planetnametwo}{Kepler-970\,b}

\newcommand{\mps}{m\,s$^{-1}$}

\newcommand{\vsini}{$v\sin{i_*}$}

\newcommand{\kepler}{{\it Kepler}}
\newcommand{\logg}{$log~g$ }

\newcommand{\um}{$\mu$m}
\newcommand{\fbol}{$F_{\mathrm{bol}}$}

\newcommand{\teff}{\ensuremath{T_{\text{eff}}}}
\newcommand\kms{km~s$^{-1}$}

\newcommand{\tess}{\textit{TESS}}
\newcommand{\ktwo}{{\textit K2}}

\newcommand{\gaia}{\textit{Gaia}}
\newcommand{\coude}{Coud{\'e}}

\shorttitle{A mini-Pleiades in the Kepler field}
\shortauthors{Barber et al.}
\graphicspath{{./}{}}

\begin{document}

\title{Transit Hunt for Young and Maturing Exoplanets (THYME) VIII: a Pleiades-age association harboring two transiting planetary systems from \kepler}

\correspondingauthor{Madyson G. Barber}
\email{madysonb@live.unc.edu}

\author[0000-0002-8399-472X]{Madyson G. Barber}%
\altaffiliation{UNC Chancellor’s Science Scholar}
\affiliation{Department of Physics and Astronomy, The University of North Carolina at Chapel Hill, Chapel Hill, NC 27599, USA} 

\author[0000-0003-3654-1602]{Andrew W. Mann}%
\affiliation{Department of Physics and Astronomy, The University of North Carolina at Chapel Hill, Chapel Hill, NC 27599, USA} 

\author[0000-0002-9446-9250]{Jonathan L. Bush}%
\affiliation{Department of Physics and Astronomy, The University of North Carolina at Chapel Hill, Chapel Hill, NC 27599, USA} 

\author[0000-0003-2053-0749]{Benjamin M. Tofflemire}
\altaffiliation{51 Pegasi b Fellow}
\affiliation{Department of Astronomy, The University of Texas at Austin, Austin, TX 78712, USA}

\author[0000-0001-9811-568X]{Adam L. Kraus}%
\affiliation{Department of Astronomy, The University of Texas at Austin, Austin, TX 78712, USA}

\author[0000-0001-9626-0613]{Daniel M. Krolikowski}%
\affiliation{Department of Astronomy, The University of Texas at Austin, Austin, TX 78712, USA}

\author[0000-0001-7246-5438]{Andrew Vanderburg}%
\affiliation{Department of Physics and Kavli Institute for Astrophysics and Space Research, Massachusetts Institute of Technology, Cambridge, MA 02139, USA}

\author[0000-0002-9641-3138]{Matthew J. Fields}
\affiliation{Department of Physics and Astronomy, The University of North Carolina at Chapel Hill, Chapel Hill, NC 27599, USA} 

\author[0000-0003-4150-841X]{Elisabeth R. Newton}%
\affiliation{Department of Physics and Astronomy, Dartmouth College, Hanover, NH 03755, USA}

\author[0000-0002-6397-6719]{Dylan A. Owens}%
\affiliation{Gemini Observatory, NSF's NOIRLab, 670 N. A'ohoku Place Hilo, Hawaii, 96720, USA}
\affiliation{Department of Physics and Astronomy, The University of North Carolina at Chapel Hill, Chapel Hill, NC 27599, USA}

\author[0000-0001-5729-6576]{Pa Chia Thao}
\altaffiliation{NSF Graduate Research Fellow}
\altaffiliation{Jack Kent Cooke Foundation Graduate Scholar}
\affiliation{Department of Physics and Astronomy, The University of North Carolina at Chapel Hill, Chapel Hill, NC 27599, USA}

\begin{abstract}
Young planets provide a window into the early stages and evolution of planetary systems. Ideal planets for such research are in coeval associations, where the parent population can precisely determine their ages. We describe a young association (\association) in the \kepler\ field, which harbors two transiting planetary systems (Kepler-1928 and Kepler-970). We identify \association\ by searching for kinematic and spatial overdensities around \kepler\ planet hosts with high levels of lithium. To determine the age and membership of \association, we combine new high-resolution spectra with archival light curves, velocities, and astrometry of stars near Kepler-1928 spatially and kinematically. We use the resulting rotation sequence, lithium levels, and color-magnitude diagram of candidate members to confirm the presence of a coeval $105\pm$10\,Myr population. \association\ may be part of the recently identified Theia 316 stream. For the two exoplanet systems, we revise the stellar and planetary parameters, taking into account the newly-determined age. Fitting the 4.5\,yr \kepler\ light curves, we find that \planetname\ is a $2.0\pm0.1R_\oplus$ planet on a 19.58-day orbit, while \planetnametwo\ is a $2.8\pm0.2R_\oplus$ planet on a 16.73-day orbit. Kepler-1928 was previously flagged as an eclipsing binary, which we rule out using radial velocities from APOGEE and statistically validate the signal as planetary in origin. Given its overlap with the \kepler\ field, \association\ is valuable for studies of spot evolution on year timescales, and both planets contribute to the growing work on transiting planets in young stellar associations.
\end{abstract}


\section{Introduction}\label{sec:intro}
Stellar clusters and associations serve as critical benchmarks for stellar and planetary astrophysics. Stars in such groups formed from the same interstellar cloud, and hence share a common (or similar) age, abundance pattern, and formation environment. The commonality makes it significantly easier to assign properties to the whole population, providing age estimates that are more precise and accurate than when the same techniques are used outside clusters \citep[e.g., Gyrochronology;][]{2007ApJ...669.1167B, vanSaders2016} and which can be applied to stars where ages are especially challenging to determine \citep[e.g., M dwarfs;][]{2021arXiv210401232K}. Such coeval associations are therefore ideal for studying how stellar and planetary properties evolve with time \citep[e.g.,][]{2019ARA&A..57..227K, 2019ApJS..245...13B,  THYMEIV}. 

Associations within the \kepler\ field have been especially valuable for stellar and planetary astrophysics. The 4.5\,yr baseline, and exquisite photometry enable precise measurements of rotation periods, even at older ages \citep[e.g.,][]{Angus2015, Aigrain2015}, providing some of the best constraints on the rotation evolution of stars past 1\,Gyr \citep{2011ApJ...733L...9M, Curtis_stall}. The four \kepler\ clusters (NGC 6866, NGC 6811, NGC 6819, and NGC 6791) have also provided a wealth of information about stellar mass-loss \citep{2012MNRAS.419.2077M}, post-main-sequence stellar evolution \citep{2012ApJ...757..190C}, and the occurrence of planets inside clusters \citep{Meibom2013}.  

The four original clusters in the \kepler\ field are all at distances of more than 1\,kpc and ages $\gtrsim$500\,Myr \citep{Batalha:2010fk}. While these older ages (compared to nearby young groups) fill an important niche in stellar spin down and post-main-sequence evolution, their distance from the Sun makes it challenging to study the low-mass members and search for small planets in the clusters.

The \ktwo\ and \tess\ mission covered many younger and more nearby clusters and star-forming regions \citep{2014SPIE.9143E..20R, VanCleve2016}. This lead to a wide range of surveys on stellar rotation \citep[e.g.,][]{2017ApJ...839...92R, Douglas:2019}, eclipsing binaries \citep[e.g.,][]{Kraus2015, David2019}, and planetary systems \citep[e.g.,][]{Gaidos:2017aa, 2019ApJ...885L..12D}. The latter included two major surveys, Zodiacal Exoplanets in Time \citep[ZEIT;][]{Mann2016a} focusing on planets in clusters observed with \ktwo, and the \tess\ Hunt for Young and Maturing Exoplanets \citep[THYME;][]{THYMEI}, focusing on similarly young stars observed with \tess. Such surveys have dramatically increased the number of known planets in young associations. However, the reliance on \ktwo\ and \tess\ data meant the overwhelming majority of stars had only 27 or 80 days of monitoring, limiting the precision and orbital period window. Associations in the \kepler\ prime field would have the full $\simeq$4.5\,yr baseline.

The availability of precise parallaxes and proper motions for millions of stars from \gaia\ \citep{Gaia_mission2016, GaiaEDR3} has enabled the discovery of new coeval stellar associations \citep[e.g.,][]{2019A&A...622L..13M, 2021arXiv210509338K}, including the 40\,Myr $\delta$ Lyr cluster that overlaps with the \kepler\ field \citep{2021arXiv211214776B}. The \texttt{FriendFinder} code\footnote{\url{https://github.com/adamkraus/Comove}} \citep{THYMEV} was designed to take advantage of \gaia\ data, by searching for potential co-moving `friends' around a user-identified young star. This method has already been useful in finding the 250\,Myr MELANGE-1 \citep{THYMEV} and MELANGE-2 (Newton et al., submitted) associations, and age-dating a planet in the Musca region of Lower-Centaurus-Crux \citep{THYMEVI}. 

To find undiscovered associations with transiting planets, we ran \texttt{FriendFinder} on \kepler\ Objects of Interest (KOI) with lithium levels indicating an age younger than Hyades \citep{2018ApJ...855..115B}. The most promising association was a group of stars nearby Kepler-1928: the candidate members showed a color-magnitude diagram (CMD) and lithium levels consistent with the Pleiades.

Here, we demonstrate that \association\ is a coeval, 105\,Myr-old group, 300\,pc from the Sun, that harbors two transiting planetary systems (\planetname\ and \planetnametwo). In Section~\ref{sec:target}, we detail our initial selection of potential members of \clustername. We describe the range of archival and new data taken for candidate members of \clustername\ in Section~\ref{sec:obs}. In Section~\ref{sec:cluster}, we show that \clustername\ is a coeval population and derive its overall properties and basic membership. Our effort to recover known and find new planets in \clustername\ is described in Section~\ref{sec:search}. This includes the identification of another member with a confirmed planet (\starnametwo). We describe the parameters of \starname\ and \starnametwo\ in Section~\ref{sec:stellar_params}, and we derive the properties of the two identified planets in the association, \planetname\ and \planetnametwo, in Section~\ref{sec:transit}. The latter signal is already confirmed, and we statistically validate the former as planetary in Section~\ref{sec:fpp}. We summarize our findings in Section~\ref{sec:summary} and briefly discuss the future utility of an association overlapping the \kepler\ field.

\section{Target Selection}\label{sec:target}

To identify known planets in previously undiscovered young associations, we ran the \texttt{FriendFinder} code \citep{THYMEV} on KOIs that appear to be young based on their lithium absorption as reported by \citet{2018ApJ...855..115B}; this initial seed list included \starname. The \texttt{FriendFinder} algorithm used \gaia\ EDR3 positions, parallaxes, and proper motions to identify stars with similar (reprojected) sky-plane tangential velocity and $XYZ$ position to a selected input source. This required an absolute radial velocity for \starname, for which we used the value from APOGEE \citep[$v_{\rm{rad}}=-26.79$\kms;][]{2020AJ....160..120J}. 

The lithium absorption levels suggested an age for \starname\ close to the Pleiades \citep{2018A&A...613A..63B}. Unbound or weakly bound associations $>$100\,Myr should have been significantly dispersed as they orbit through the Galaxy \citep{2019ARA&A..57..227K}, so we used a generous selection. We included any star with a parallax uncertainty $\sigma_{plx}<0.5$\,mas, a tangential velocity $v_{\rm{tan}}<5$\kms\ from \starname\ and a physical separation $\rm{S}<50$\,pc from \starname. This yielded 1007 candidates. 

Our selection assumed the group is circular and centered around \starname, neither of which is likely to be true. However, our aim was not to make a perfectly clean list or a complete list of members. Rather, we aim to select a generous list that contains enough members to derive an age for the planet hosts. If the population extends beyond the edge of our search region, a future census can aim for greater spatial completeness. 

\begin{figure}[tb]
    \centering
    \includegraphics[width=0.48\textwidth]{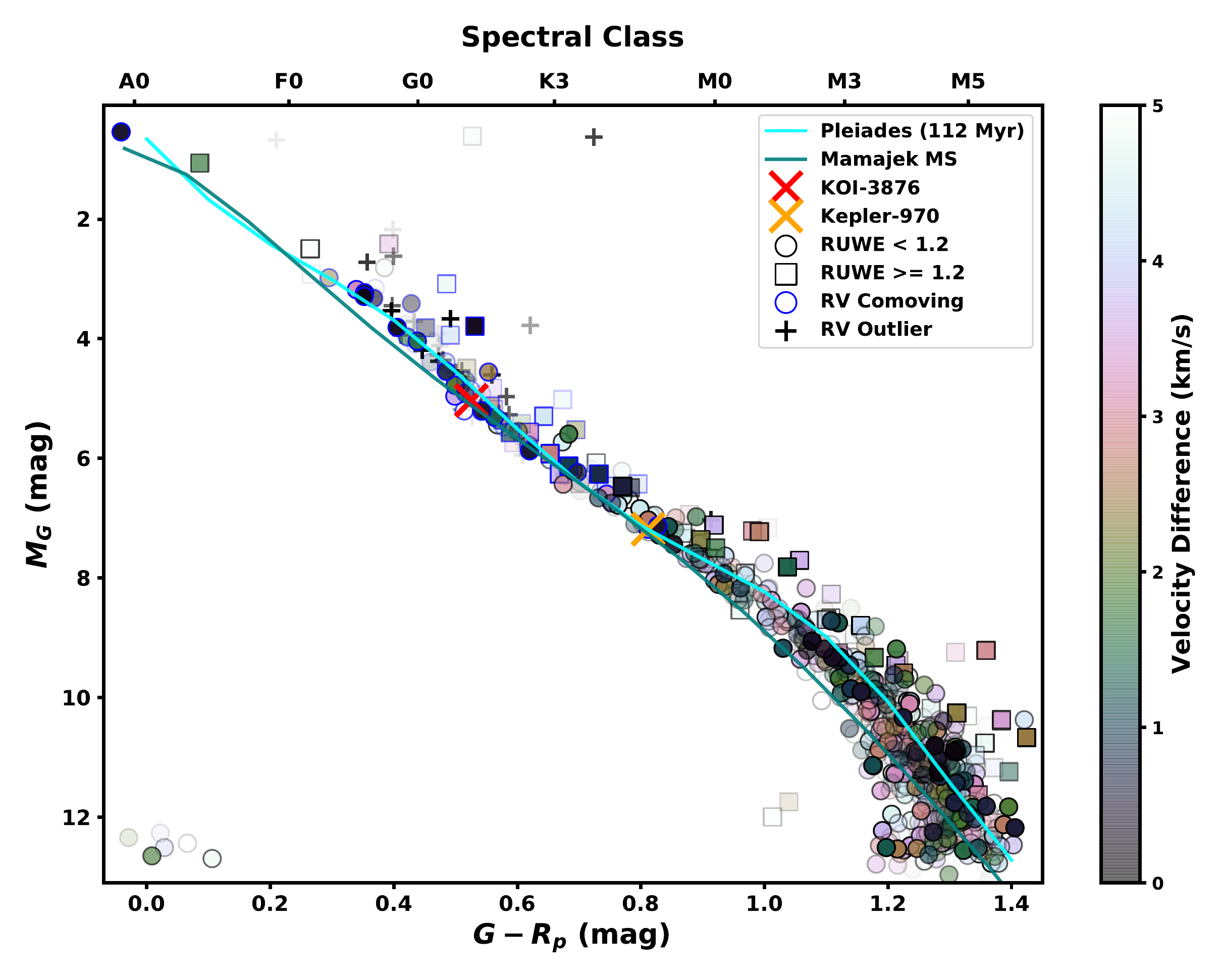}
    \caption{\gaia\ color-magnitude diagram of all stars within 5\,\kms\ in tangential velocity and 50\,pc of \starname, excluding those with high contaminated photometry \citep{EDR3photometry}. Points are color-coded by the difference between their expected and observed tangential velocity assuming a perfect $UVW$ match to \starname\ and assigned transparency based on their three-dimensional distance from \starname. Stars with radial velocities consistent with \starname\ are marked with blue circles and those with discrepant velocities are changed to plus signs (and excluded from the color-coding). Approximate main- and Pleiades- single-star sequences are shown as colored lines \citep{2013ApJS..208....9P, 2019A&A...628A..66L}.  }
    \label{fig:CMD}
\end{figure}

The spread of the candidate member color-magnitude diagram (CMD) indicated significant contamination from field stars (Figure~\ref{fig:CMD}). However, the CMD also harbors a sequence of the closest stars (in tangential velocity) consistent with the Pleiades single-star sequence, matching the age suggested by the Li levels in the spectrum of \starname.

\section{Observations}\label{sec:obs}

\subsection{Optical spectra from McDonald 2.7 m \coude}\label{sec:tull}

We observed \name\ and 22 association candidates (Section \ref{sec:target}) with the \coude\ spectrograph on the Harlan J. Smith 2.7m telescope at the McDonald Observatory. The Robert G. Tull \coude\ is a cross-dispersed echelle spectrograph, delivering a $R\sim60,000$ spectral resolution from 3400--10000 \AA\ using the 1$\farcs$2 slit \citep{Tull1995}. Observations were taken over three nights from two observing runs, on 2021 July 9 and 2021 August 26 and 27. 

The sample was selected to include association candidates that could be observed with the \coude\ in modest exposure times ($G<13$), and spectral types later than mid-F ($B_p-R_P > 0.55$), where we expect lithium absorption to be a sensitive age diagnostic. These cuts yielded 78 stars of the initial 1007. While observing, we prioritized targets with $v_{\rm{tan}}<$ 4\kms\ from \starname\ and observability (e.g., favorable airmass). 

We reduced all spectra with a custom python implementation of the standard IRAF procedures. Wavelength calibration made use of ThAr lamp spectra taken at the beginning, middle, and end of each night. The signal-to-noise of our spectra ranged between 14 and 60 per resolution element. 

To assess whether association candidates are co-moving in three dimensions, we measure radial velocities using spectral-line broadening functions (BFs). The BF is a linear inversion of an observed spectrum with a narrow-lined template and represents the average stellar absorption-line profile. This profile (the BF) can be fitted with a rotationally-broadened line profile to measure the stellar radial velocity and \vsini. We compute BFs for 34 spectral orders between 4300 and 9800 \AA\ that is free of telluric contamination using the {\tt saphires} python package \citep{Tofflemireetal2019}. BFs from individual orders are combined into a single, high SNR BF and fit with a rotationally broadened profile \citep{Gray1992}. Narrow-lined templates, specific to each star, are taken from the \citet{2013A&A...553A...6H} PHOENIX model suite at the \teff\ closest to that provided by the \tess\ Input Catalog \citep[v8.0;][]{TIC2019}. Radial-velocity errors depend on the S/N and rotational broadening but are generally on the order of 0.1 \kms. Measurements from the \coude\ spectra are provided in Table \ref{tab:fullsample}.

\subsection{Archival Velocities}\label{sec:archRVs}
In order of preference, we drew radial velocities for candidate \association\ members from our spectra (Section~\ref{sec:tull}), the second \gaia\ data release \citep[DR2;][]{DR2_velocities}, the sixteenth APOGEE data release \citep[DR16; ][]{2020AJ....160..120J}, and the fifth LAMOST data release \cite[DR5; ][]{2015RAA....15.1095L, 2019yCat.5164....0L}. In the instance where a star had multiple velocities from the same source, we used the weighted mean and error. We did not combine multiple velocities from different sources. We applied an offset to the LAMOST velocities of +4.54 \kms\ based on the comparison from \citet{2018A&A...620A..76A}. There may be additional zero-point differences between the velocity sources, but these are likely smaller than the internal velocity spread within the group. 

In total, we adopted \gaia\ RVs for 56 stars, APOGEE RVs for 5 stars, and LAMOST RVs for 25 stars. This was in addition to velocities from our Coude spectra for 22 candidate member stars and \starname. The adopted velocities are given in Table~\ref{tab:fullsample}.

\subsection{Kepler and TESS light curves}\label{sec:lc}

\subsubsection{Light curves for transit search and characterization}

We searched for \kepler\ photometry for all 1007 candidates within the Mikulski Archive for Space Telescopes (MAST). A total of 84 targets had \kepler\ data, the majority of which had data from all quarters (Q0-Q17). We restricted our analysis to long-cadence data (30\,m), as short-cadence was not available for any of the candidate planet hosts and short cadence data is not needed for our transit search or rotation estimates.

Where available, we used the Pre-search Data Conditioning Simple Aperture Photometry \citep[PDCSAP;][]{Stumpe2012, Smith2012}. This included \starname\ in both \kepler\ and \tess\ data, all candidate members with \kepler\ light curves, and the 15 targets with 2-minute cadence photometry from \tess. For the 41 with 30-minute cadence TESS data, we used the Quick-Look Pipeline light curves \citep{2020RNAAS...4..204H} for our planet search. Table~\ref{tab:fullsample} lists which stars have \kepler\ and/or \tess\ photometry that was used for our planet search. More details on our transit search can be found in Section~\ref{sec:search}.

The remaining 885 sources had no pre-extracted \tess\ or \kepler\ light curves. We did not extract additional curves from the full-frame \tess\ images for our planet search or characterization. The association is more than 300\,pc away; most of the remaining stars were too faint \citep[QLP limit of $T<13.5$][]{2020RNAAS...4..204H}, to extract a light curve precise enough for our planet search. However, many such systems are still useful for measuring rotation periods, as we discuss in the next section.

\subsubsection{Stellar rotation periods}\label{sub:rot_collection}

To assess the membership and age of the candidate association we collected stellar rotation periods from the literature, which we supplemented with our measurements from \kepler\ and \tess\ light curves. First, candidate members were cross-matched against \citet{Nielsen:2013}, \citet{McQuillan2013, 2014ApJS..211...24M}, and \citet{Santos2019, Santos2021} to identify literature rotation periods. We matched candidate members to the \kepler\ Input Catalog \citep[KIC;][]{Brown2011} using the \texttt{astroquery} package, selecting the KIC match by smallest on-sky separation. If a KIC ID was not returned from this search, we manually identified KIC IDs using the \kepler\ Target Search portal. We identified $56$ candidate members with available literature rotations, all but five of which have rotation measurements in multiple catalogs. For candidates that appear in only one catalog, we adopt the single measurement value. In cases where a star had measurements from more than one source, we adopt the average of the measurements as the rotation period. Only one object, KIC 3743810, had a conflicting rotation period between catalog sources. Based on a visual examination of this object's \kepler\ PDCSAP light curve, we selected the value from \citet{Nielsen:2013}.

For stars without literature rotation periods, we performed our analysis on \kepler\ and \tess\ data. Priority was given to \kepler\ PDCSAP data followed by \tess\ full-frame images (FFI). For each star with available \kepler\ data, we searched each single-quarter \kepler\ light curve for rotation periods between $0.1-50$ days using the Lomb-Scargle algorithm \citep{LombScargle}. We selected the initial rotation from the quarter returning the rotation period with the highest periodogram power. To confirm these measurements, we phase-folded the single-quarter light curves to the candidate period and examined the signals' consistency across quarters. We performed an eye-check in the style of \citet{2021ApJ...921..167R}, labeling obvious rotations as Q0, questionable rotations as Q1, spurious detections as Q2, and non-detections as Q3.

In total, we identified usable rotations of quality Q0 or Q1 in $11$ out of 15 of the stars with \kepler\ data and no literature rotation period. As a check, we ran 50 stars with literature rotation periods using our own analysis. Of these 50, 47 were recovered within 10\% of the literature measurement, 2 within 20\%, and our last measurement was a half alias of the literature measurement. We considered this agreement to be excellent.

For the rest of the 933 candidates without rotations found in the literature or through our \kepler\ light curve measurements, we searched for signatures of rotation in CPM light curves extracted from the \tess\ Full Frame Image data. We did not use the QLP curves, as we found they did not preserve the stellar rotation signal reliably. Instead, we generated \tess\ light curves from the FFI cutouts. We first created raw flux light curves from the FFI cutouts centered on each candidate. Then, we generated a Causal Pixel Model (CPM) of the telescope systematics using the \texttt{unpopular} package \citep{2021arXiv210615063H} for each individual star. 

Before the raw light curve was computed, we background subtracted using the median value of the 300 dimmest pixels on the cutout. For running \texttt{unpopular}, we used a one pixel aperture centered on the target. We used a 5x5 exclusion region around the target before selecting 100 predictive pixels using the ``Similar Brightness'' method. We split the light curve into 100 contiguous sections for predicting and testing, with an L2 regularization value of 0.1. 

We subtracted the resulting CPM systematics from the initial light curves. For 26 targets, we failed to extract a usable CPM curve; 8 because of no clear matching TIC ID and 18 because the CPM extraction process failed. After searching each single-sector light curve of each star for rotation periods from $0.1-30$ days using the Lomb-Scargle algorithm, we repeated the same rotation selection and quality check procedure as outlined for the \kepler\ data. We found $64$ quality Q0 or Q1 rotations from the 907 \tess\ CPM-subtracted light curves available. 

In total, we were able to assign rotation periods to $131$ candidate members, all of which are reported in Table~\ref{tab:fullsample}; 67 periods were determined based on \kepler\ data, 11 newly calculated and 56 from literature, and 64 were calculated from \tess\ data.

Based on variations in the extracted rotation period between \tess\ sectors and/or \kepler\ quarters, we estimate rotation period errors to be $\simeq$10\% for our measurements. This is larger than the expected errors just considering signal-to-noise and Lomb-Scargle errors from bootstrapping, likely due to differential rotation and spots appearing and disappearing on the surface of the star \citep{2021ApJ...921..167R}.

\subsection{Archival photometry and astrometry}
We download positions, parallaxes, proper motions, and $B_P$, $R_P$ and $G$ photometry for all candidate members of \association\ using the third \gaia\ Early Data Release \citep[EDR3;][]{GaiaEDR3}. For \starname\ and \starnametwo\ (see Section~\ref{sec:search}), we also retrieved photometry from the Two-Micron All-Sky Survey \citep[2MASS;][]{Skrutskie2006}, the Wide-field Infrared Survey Explorer \citep[WISE; ][]{allwise}, and the AAVSO All-Sky Photometric Survey \citep[APASS; ][]{apass}. Photometry for \starname\ and \starnametwo\ are listed in Table~\ref{tab:prop}.

\section{The \association\ Association}\label{sec:cluster}

Our goal for the rest of this section is to demonstrate that the friends of \starname\ represent a coeval population and to determine the age and kinematic properties of the association. We refer to this group as \association\ (Membership and Evolution by Leveraging Adjacent Neighbors in a Genuine Ensemble), following the naming convention in \citet{THYMEV}, although it is likely that \association\ is a component of the larger Theia 316 string (see Section~\ref{sec:theia}).

As we detail below, we found that \association\ members closely match those of the Pleiades in Lithium equivalent widths (Section~\ref{sec:lithium}), rotation (Section~\ref{sec:rotation}), and CMD. We combined these to estimate that \association\ is $105\pm10$\,Myr. From a subset of high-probability members, we derived the Galactic position (XYZ) and kinematic (UVW) parameters (Section~\ref{sec:kinametics}).
Using stellar rotation periods and radial velocities, we estimated that 50-60\% of the original \association\ candidate list are real members (Section~\ref{sec:contam}). 

\subsection{Radial Velocities}\label{sec:RVs}

\begin{figure}[b]
    \centering
    \includegraphics[width=0.47\textwidth]{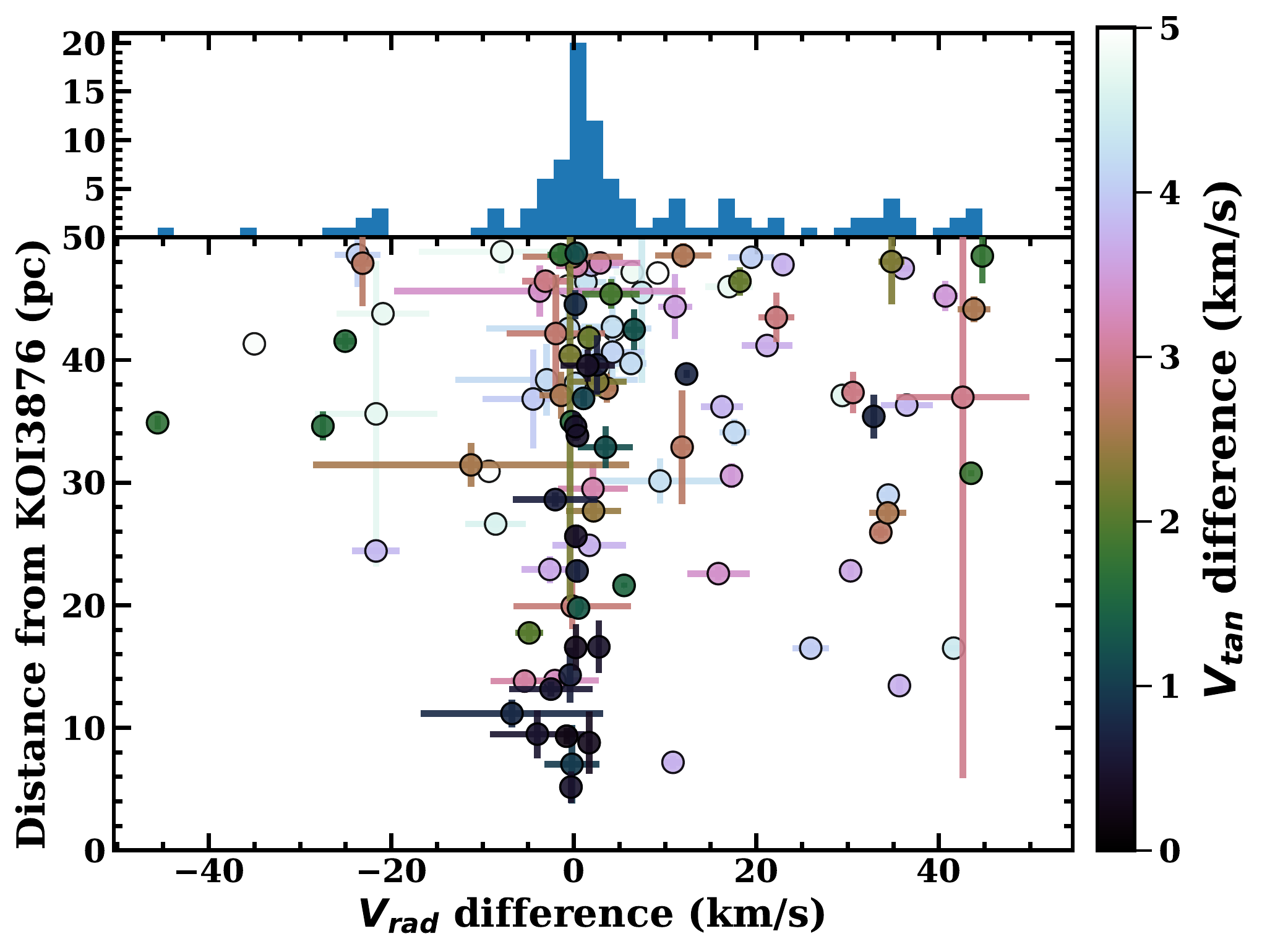}
    \caption{Three-dimensional distance from \starname\ as a function of the difference between the observed and predicted radial velocity of candidate members of \association, color-coded by the tangential velocity offset from \starname. Errors are shown, although many stars have distance and velocity errors comparable to or smaller than the point size. The top panel shows a histogram of the radial velocity difference. The predicted velocities were calculated assuming each candidate has an identical $UVW$ to \starname. We do not expect all true members to be exactly consistent with zero velocity difference: \starname\ is not necessarily at the kinematic center of the group, the group has some intrinsic velocity spread, and some stars may be tight binaries. The over-density of points near-zero velocity difference is strong evidence that the association is real and extends to at least to 50\,pc from \starname. 
    \label{fig:groupRV}
    }
\end{figure}

\texttt{FriendFinder} generated a predicted radial velocity for each candidate member under the assumption that every member star has an identical $UVW$ to \starname\ (i.e., re-projecting the $UVW$ of \starname\ to the position of each star). Since radial velocities were not used in the target selection (only $XYZ$ and tangential velocities), the difference between the predicted and measured velocities can be used to test if the stars are truly co-moving. We show this comparison for candidate members of \association\ in Figure~\ref{fig:groupRV}, taking into consideration the three-dimensional and tangential velocity offsets. 

The radial velocities of candidate members are heavily clustered within $\simeq2$ \kms\ of the predicted values. The over-density is also highest for stars closer to \starname\ in tangential velocity and position. Based on the Besan\c con Galactic model \citep{2014A&A...564A.102C}, typical thin disk stars with similar $XYZ$ to \starname\ will have a velocity spread of $\simeq40$\kms\ and centered more than 30\kms\ from the locus seen in Figure~\ref{fig:groupRV}. Thus, the probability of such a buildup by chance is negligibly small. Interestingly, there is still an overdensity of stars with consistent radial velocities even 50\,pc from \starname. This suggests that the true population is much larger.

The distribution in Figure~\ref{fig:groupRV} also highlights a challenge: there is a large population of stars with radial velocities well outside the central locus. Such contaminants include stars near \starname\ and \starnametwo\ in position and tangential velocity. Further, the overdensity in velocity space is not necessarily coeval; overdensities in kinematic space can occur for other reasons, such as dynamical perturbations from the Galactic bar \citep{2010ApJ...717..617B}. 

\subsection{Metallicity}\label{sec:metallicity}

We estimated the iron abundance ([Fe/H]) of \association\ from the literature metallicities of candidate members. After excluding targets with radial velocities more than 10\kms\ from \starname, we were left with [Fe/H] estimates for 27 candidate members from LAMOST \citep[18 stars,][]{2019yCat.5164....0L}, APOGEE \citep[7 stars,][]{2020AJ....160..120J}, and prior analyses of the two planet hosts discussed further in Section~\ref{sec:stellar_params}. We excluded three of these stars, which had anomalously low iron abundances ([Fe/H]$<-0.3$) suggesting they were non-members. The remaining 24 had a mean iron abundance of [Fe/H] = 0.04+/-0.03. The weighted mean and error gave a similar result and smaller error, but likely underestimates the role of systematics common to all determinations from a single source as well as possible offsets between the scale used by LAMOST, APOGEE, and the measurements for the two planet hosts. We adopted the simple mean and standard error as the group iron abundance. 

\subsection{Lithium}\label{sec:lithium}

Lithium is quickly burned in the cores of stars. As a result, surface lithium is slowly depleted over time in low-mass stars, yielding a mass-dependent lithium sequence that shifts with age \citep[e.g.,][]{2017MNRAS.464.1456J, 2017AJ....153..128C}. We can therefore confirm that \association\ is a coeval population and estimate its age by comparing the lithium levels as a function of color to measurements from other known clusters. 

We estimated the equivalent width (EW) of the Li 6708\,\AA\ line for 23 stars (including \name) using the Coude spectra described in Section~\ref{sec:tull}. Using our measured radial and rotation velocities from the BF analysis, we shifted each spectrum to zero velocity and compared it to a rotationally broadened template of the same \teff. We then interactively defined regions of line-free continuum between 6685 and 6730 \AA, and the bounds of the EW integration. The main sources of uncertainty in the EW integration come from the continuum level and the bounds of integration. We attempted to account for these two factors using a bootstrap approach. First, we fit the interactively defined continuum regions with a linear slope, which is appropriate given the spectral window considered, using {\tt emcee} with uniform priors on the slope and y-intercept and assuming Gaussian errors.

The fit was at least 50 times the the chain auto-correlation time, sufficient for convergence. Five auto-correlation times are removed as burn-in from the final posterior. From these posteriors, 1000 random draws from the slope and y-intercept were used to normalize the spectrum; for each realization, the Li absorption line was numerically integrated 10 times where the integration bounds were varied randomly from a normal distribution with the width of a resolution element. This procedure resulted in 10,000 EW[Li] measurements, and we took the median and standard deviation as our final measurement and its uncertainty, respectively. 

Past detections of Li with the same observational setup and our typical spectrum SNR indicate we were sensitive to Li down to equivalent widths of 20m\AA\ or better \citep{THYMEV, THYMEVI}, so we report this as our upper limit when no line is detected. One star (Gaia EDR3 2052858307226740352) had a \vsini$>50$\kms, which made the extraction of the Li line unreliable. So, we instead reported a $<70$m\AA\ upper limit for this source based on earlier detections on similarly broadened spectra. 

\begin{figure}[!t]
    \centering
    \includegraphics[width=0.47\textwidth]{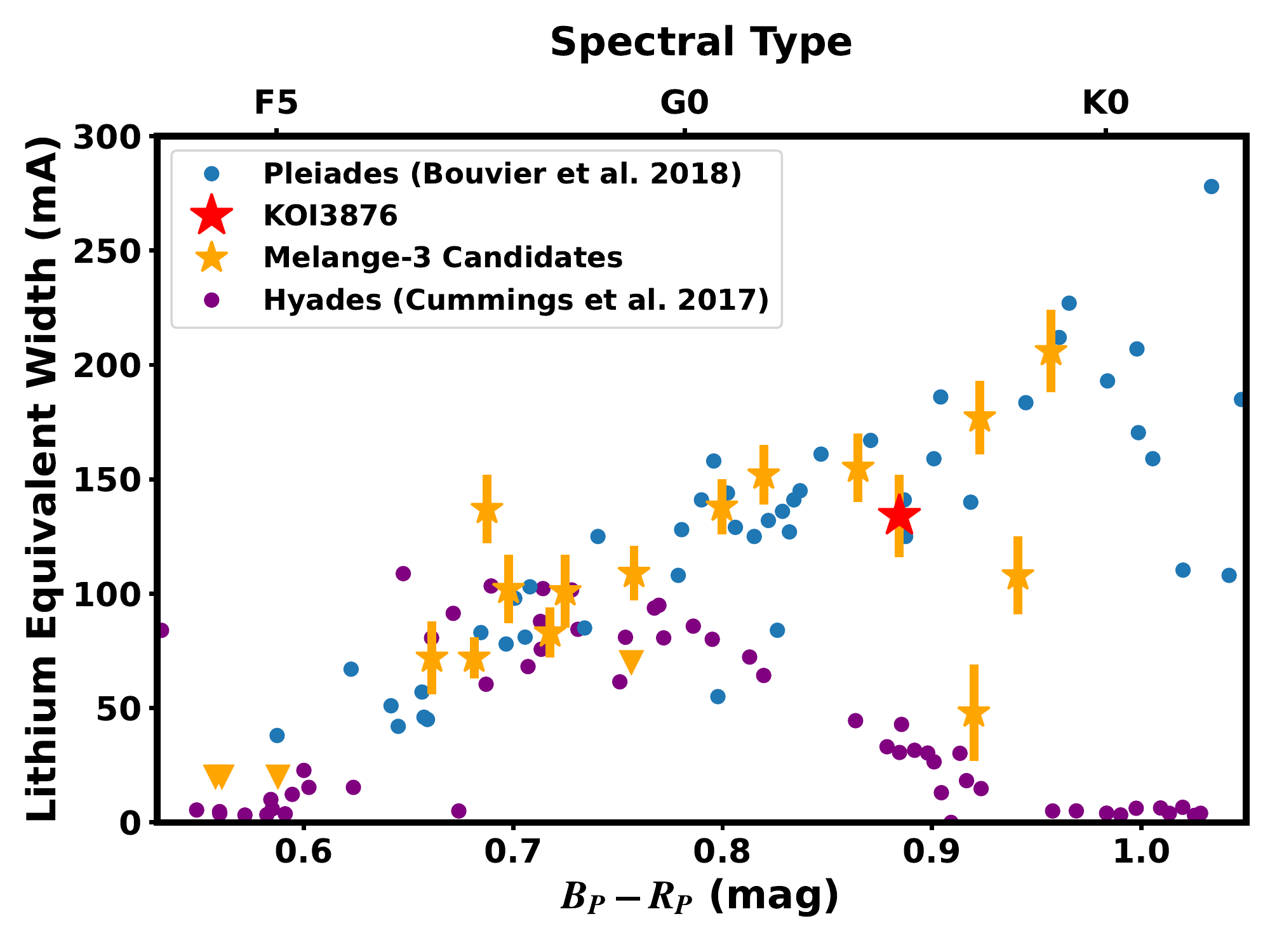}
    \caption{Lithium equivalent width as a function of \gaia\ $B_P-R_P$ color for candidate friends of \starname\ (\association; orange), \starname\ (red), and members of the 125\,Myr Pleiades from \citet{2018A&A...613A..63B} and $\simeq$700\,Myr Hyades \citep{2017AJ....153..128C}. Triangles indicate upper limits. We have excluded \association\ candidates with velocities inconsistent with membership (as defined in Section~\ref{sec:contam}). The \association\ sequence is consistent with that from Pleiades members; only one star has an anomalously low Li (Gaia EDR3 2044150037698600448) compared to the Pleiades sequence. The high levels of Lithium seen in the mid-G dwarfs alone demonstrate that \association\ is much younger than Hyades. \starnametwo\ has weak lithium (EqW$<40$m\AA), but at $B_P-R_P=1.5$, we expect it to be depleted at this age; we left it off the edge of the figure.
    \label{fig:lithium}
    }
\end{figure}

For \starname, we estimated a EW[Li]$=134\pm18$m\AA. This is marginally higher than (but consistent with) the value from \citet{2018ApJ...855..115B} (EW[Li]$=120\pm7$m\AA). We attribute this difference to \citet{2018ApJ...855..115B}'s removal of the Fe line at 6707.44\AA. We did not attempt to correct for this contamination or from broad molecular contamination in the cooler stars. Fe line contamination sets a (systematic) limit on the precision of our equivalent widths at the $\simeq$10\% level, comparable to the measurement errors. The difference was small compared to the offset in Li levels between clusters. We used our Li measurements for all targets for consistency.

Two spectra (Gaia EDR3 2101333021814076800 and 2048317736525727488) had two clear sets of lines, indicating double-lined spectroscopic binaries (SB2s). For our Li measurements, we measured each line individually with a manually-applied velocity offset and combined the two equivalent widths. 

In Figure~\ref{fig:lithium}, we compared the Li sequence for \association\ to that from the $\simeq$112\,Myr Pleiades \citep{2018A&A...613A..63B} and the 650-700\,Myr Hyades \citep{2017AJ....153..128C}. The \association\ sequence is nearly identical to that from the Pleiades. The Li sequences of nearby clusters from \texttt{BAFFLES} \citep{BAFFLES2020} suggested an age between 85\,Myr and 200\,Myr This age range is conservative, as the bounds can only be set using the set of clusters with ages and extant lithium sequence measurements. 

\subsection{Rotation}\label{sec:rotation}
\begin{figure}[tbhp]
    \centering
    \includegraphics[width=0.47\textwidth]{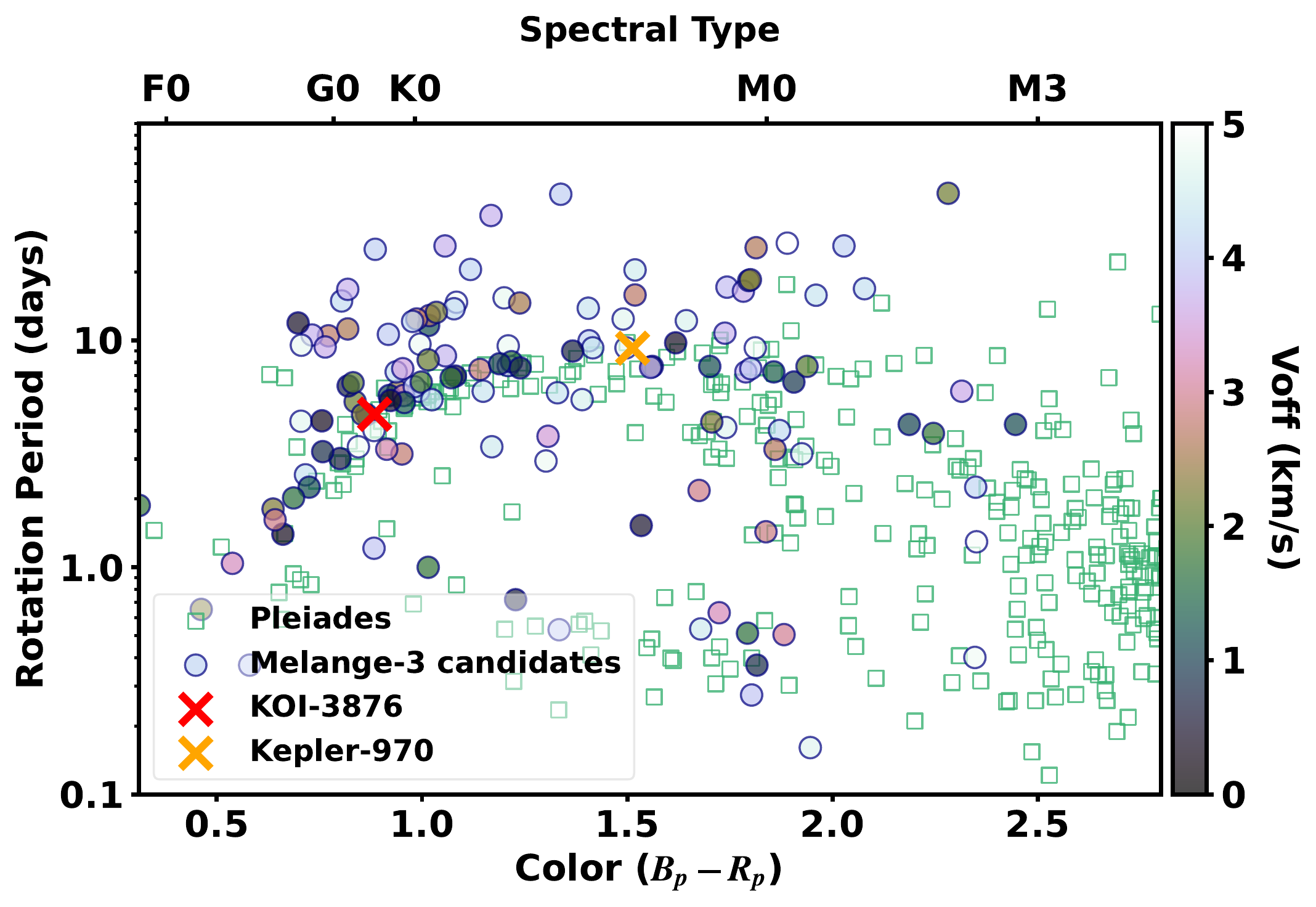}
    \caption{Rotation periods of candidate members of \association\ (dark circles) as a function of \gaia\ $B_P-R_P$ color. Only stars with literature measurements or Q0 or Q1 rotations are shown. For reference, we show rotation periods from $\simeq$110\,Myr Pleiades \citep[green squares;][]{Rebull2016}. Stars are color-coded by their tangential velocity difference compared to \starname. The stars on the Pleiades gyrochrone are also generally closer to \starname\ in tangential velocity. Because of the distance to the cluster, we have few rotation periods past M1. 
    \label{fig:rotation}
    }
\end{figure}

We used the rotation periods from Section~\ref{sub:rot_collection} to better assess the age and membership of \association. Coeval members should follow a rotation sequence in color \citep[a gyrochrone;][]{Barnes2003, 2007ApJ...669.1167B}, which we show in Figure~\ref{fig:rotation}. The distribution is consistent with that from the Pleiades, further validating the Li-based age. 

To determine the number of stars with rotation periods consistent with the Pleiades, we used a three-step cut. The first cut required stars with a $B_P-R_P<0.7$ to have a rotation period $\leq3$ days. The second cut required stars with $0.7\leq B_P-R_P<1.0$ to have a rotation period $\leq7.5$ days. The third cut required stars with $B_P-R_P\geq1.0$ to have a rotation period $\leq10$ days. We note that these cuts are somewhat qualitative but designed to capture $>95$\% of the Pleiades rotators (see Figure~\ref{fig:rotation}) and remove most field stars. These cuts yielded 92 stars consistent with the Pleiades rotation sequence.  Most of the slower rotators are likely to be field interlopers, as they are (statistically) further from \starname\ in both three-dimensional distance and tangential velocity. \association\ stars rotating faster than the Pleiades slow rotator sequence are considered association members, as the Pleiades contains fast rotators and many of the fast rotators may be binary members \citep{2016AJ....152..114R,Douglas:2019}.

Of 1007 initial candidates, finding Pleiades-like rotation periods for just 92 stars initially appeared to be an unexpectedly low success rate. However, for the overwhelming majority of the 876 stars with a light curve but no rotation period, no period could have been measured even if one was present (mostly due to intrinsic faintness). For example, of the 925 stars with a matching TIC ID, but no rotation period from \kepler\ data, 751 were either too faint ($T\gtrsim15$) or too contaminated by nearby stars to extract a usable CPM curve. Thus, the difference is mostly due to how much deeper \gaia\ can retrieve precise astrometry for stars far fainter than for which \tess\ can provide rotation periods.

\subsection{Isochronal age}\label{sec:isochrone}

We estimated the age of \association\ by comparing the CMD to the PARSEC (v1.2S) models \citep{PARSEC}. We used a mixture model, as detailed in \citet{THYMEVI}\footnote{\url{https://github.com/awmann/mixtureages}}, based on the method outlined in \citet{HoggRecipes}, and wrapped in the \texttt{emcee} Markov-Chain Monte-Carlo sampling algorithm \citep{Foreman-Mackey2013}. To briefly summarize, we fit the population with the combination of two models. The first described the single-star member sequence drawn from PARSEC models. The second is an outlier population, which may contain a mix of populations (e.g., binaries, field interlopers, and stars with erroneous parallaxes or photometry). The fit included six free parameters: the association age ($\tau$), the average reddening across the association ($E(B-V)$), the amplitude of the outlier population ($P_B$), the offset of the outlier population from the main population CMD ($Y_B$ [mags]), the variance of the outliers around the mean ($V_B$ [mags$^2$]), and a term to capture missing uncertainties or differential reddening across the association ($f$ [mags]). The full likelihood function can be found in the Appendix of \citet{THYMEVI}.

Reddening was limited to $<0.2$ mag based on the three-dimensional extinction map from \citet{2019ApJ...887...93G}. To ensure uniform sampling in $\tau$, we re-sampled the model grid in equal steps around the expected age (50-300\,Myr). All other parameters also evolved under uniform priors, bounded only by physical or practical limits ($0<P_B<1$, $-10<Y_B<10$, $0<V_B<10$, $0<f<2$).

\gaia\ photometry was available for all candidate members and was generally far more precise than other available photometry. Many stars were also resolved as binaries (or the target and a background star) in \gaia, but seen as a single source in 2MASS and KIC photometry \citep{Brown2011}. We therefore restricted our analysis to \gaia\ magnitudes. 

While the mixture model can handle high contamination rates by making $P_B$ larger, tests suggest that when $P_B\gtrsim0.4$, the model often calls stars in the main population outliers, instead of fitting the background population as the true one (yielding a field age). As we show in Section~\ref{sec:contam}, the background contamination rate is likely $\gtrsim0.4$. Further, for the isochrone analysis, there will be additional true members that should be treated as outliers due to binarity, poor photometry, or poor parallaxes. As a result, in addition to limiting the sources to objects within the colors and magnitudes bounded by the model grid, we applied the following restrictions to the input data:
\begin{itemize}
    \item a renormalised Unit Weight Error \citep[RUWE; ][]{GaiaEDR3} $<1.3$,
    \item $1.0+0.015(B_P-R_P)^2<$ \texttt{phot\_bp\_rp\_excess\_factor} $< 1.3 + 0.06*(B_P-R_P)^2$,
    \item physical separation within 40\,pc of \starname,
    \item tangential velocity within 4\,\kms\ of \starname, and
    \item a radial velocity within $3\sigma$ of the group or no radial velocity.
    \end{itemize}

These cuts were designed to mitigate, but not completely remove, issues of binarity, data quality, non-member interlopers, as well as limitations of the models. The mixture model approach means they do not need to completely remove bad data or non-members. Taken individually, none of these had a significant impact on the derived age except the \texttt{phot\_bp\_rp\_excess\_factor} criteria. Stars with flux contamination tended to have CMD positions far {\it below} the main-sequence, likely due to poor or non-detections in $B_P$ reported as detections near the limits \citep{GaiaEDR3Validation}. After all cuts were made, we were left with 205 stars.

\begin{figure}[tbhp]
    \centering
    \includegraphics[width=0.48\textwidth]{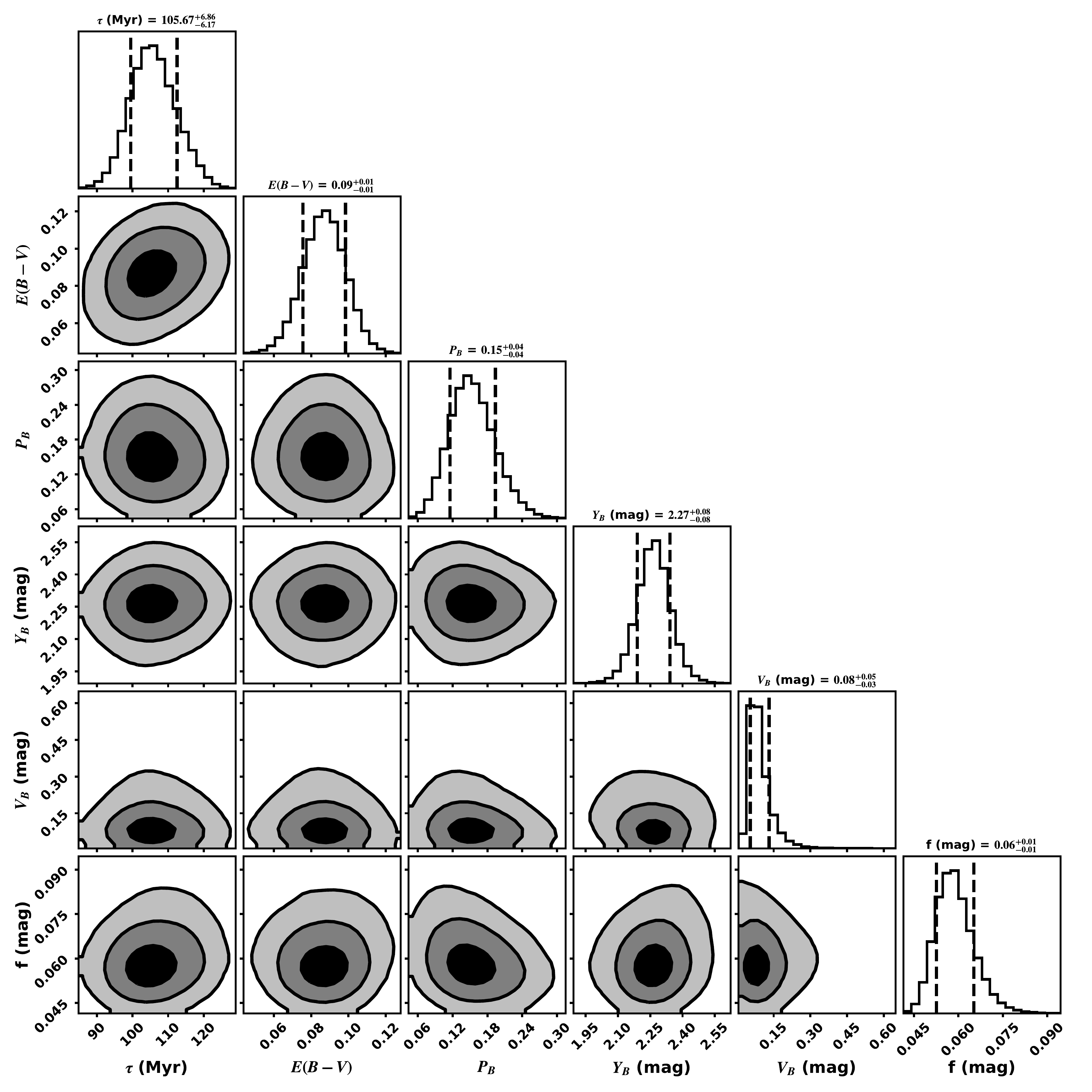}
    \includegraphics[width=0.48\textwidth]{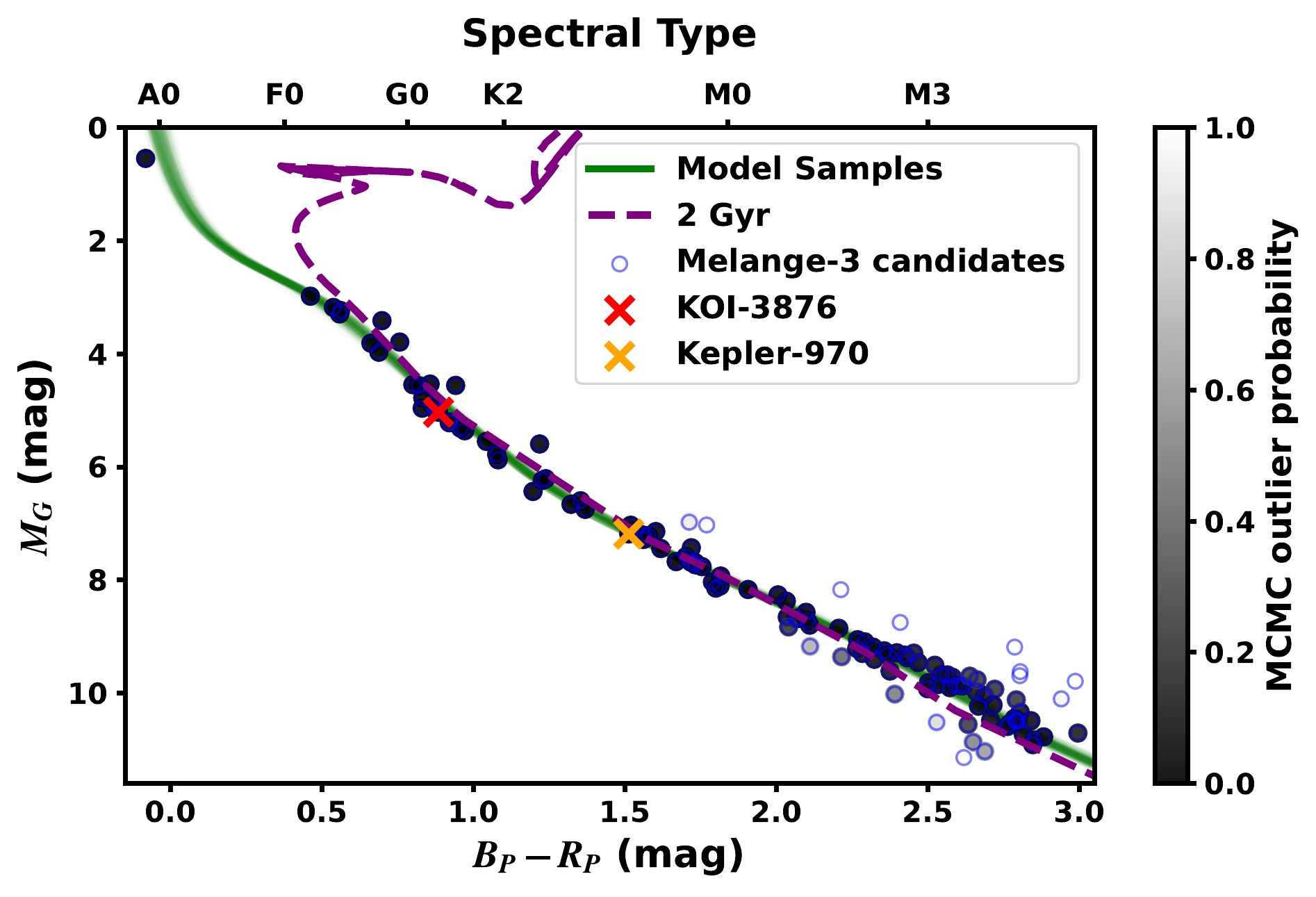}
    \caption{Comparison of the PARSEC model isochrones to candidate members of \association. The top shows the corner plot of our MCMC mixture model comparison, with contours corresponding to $1\sigma$, $2\sigma$, and $3\sigma$. The bottom plot shows the \gaia\ $G$ versus $B_P-R_P$ CMD of stars included in the MCMC (blue circles). Each point is shaded based on its average outlier probability as determined by the relative strength of the two mixture components. Most of those flagged as outliers are likely to be non-members or binaries that evaded our selection cuts. \starname\ and \starnametwo\ are shown as red and orange stars. Both have low ($<1\%$) outlier probabilities. The green lines are 200 PARSEC isochrones with parameters drawn (randomly) from the MCMC posterior. The purple dashed line indicates a `field' (2 Gyr) population, highlighting that a lot of the age leverage comes from the pre-main-sequence mid-M dwarfs and the small number of AF stars. }
    \label{fig:isochrone}
\end{figure} 

As we show in Figure~\ref{fig:isochrone}, the resulting fit yielded an age of 106$^{+7}_{-6}$\,Myr. As a test of the systematic errors, we ran the same fit using models from the Dartmouth Stellar Evolution Program \citep[DSEP,][]{Dotter2008}. Using the DSEP models with magnetic enhancement described in \citet{Feiden2012b} gave a similar age of 101$\pm$7\,Myr. DSEP models without magnetic enhancement gave a younger age of 96$\pm$10\,Myr, although a poorer fit to the coolest stars in the sample. Using (PARSEC) non-Solar metallicity models changed the age at the $5$\,Myr level, but favoring older ages. All ages agree with each other at the 1-2$\sigma$ level. These ages are also consistent with our measurements using lithium and stellar rotation (Sections~\ref{sec:rotation} and \ref{sec:lithium}), both of which indicated an age close to Pleiades \citep[$\simeq$112\,Myr; ][]{Dahm2015}. 

\subsection{Galactic Position and Kinematics}\label{sec:kinametics}

\begin{figure*}[tbhp]
    \centering
    \includegraphics[trim=90 0 150 0,clip=true,width=0.98\textwidth]{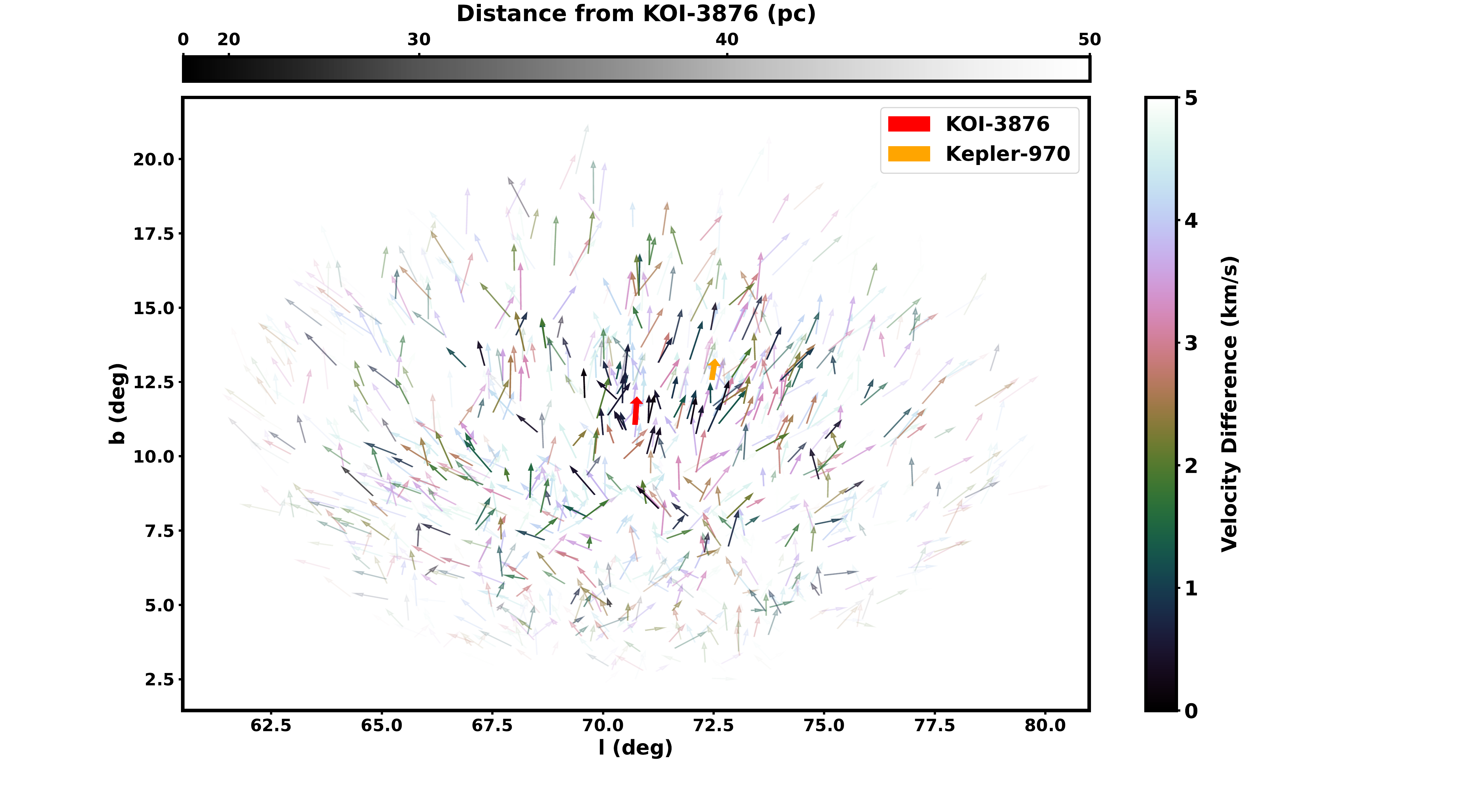}
    \caption{Galactic coordinates and motions for all 1007 candidate members of \association. The red arrow shows \starname, Kepler-970 is shown in orange, and other stars are color-coded by their tangential velocity offset from \starname\ with transparency set by their physical distance from \starname. Arrows indicate the direction and (relative) magnitude of the proper motion. 
    }
    \label{fig:ProperMotion}
\end{figure*}

\begin{figure*}[p]
    \centering
    \includegraphics[width=0.99\textwidth]{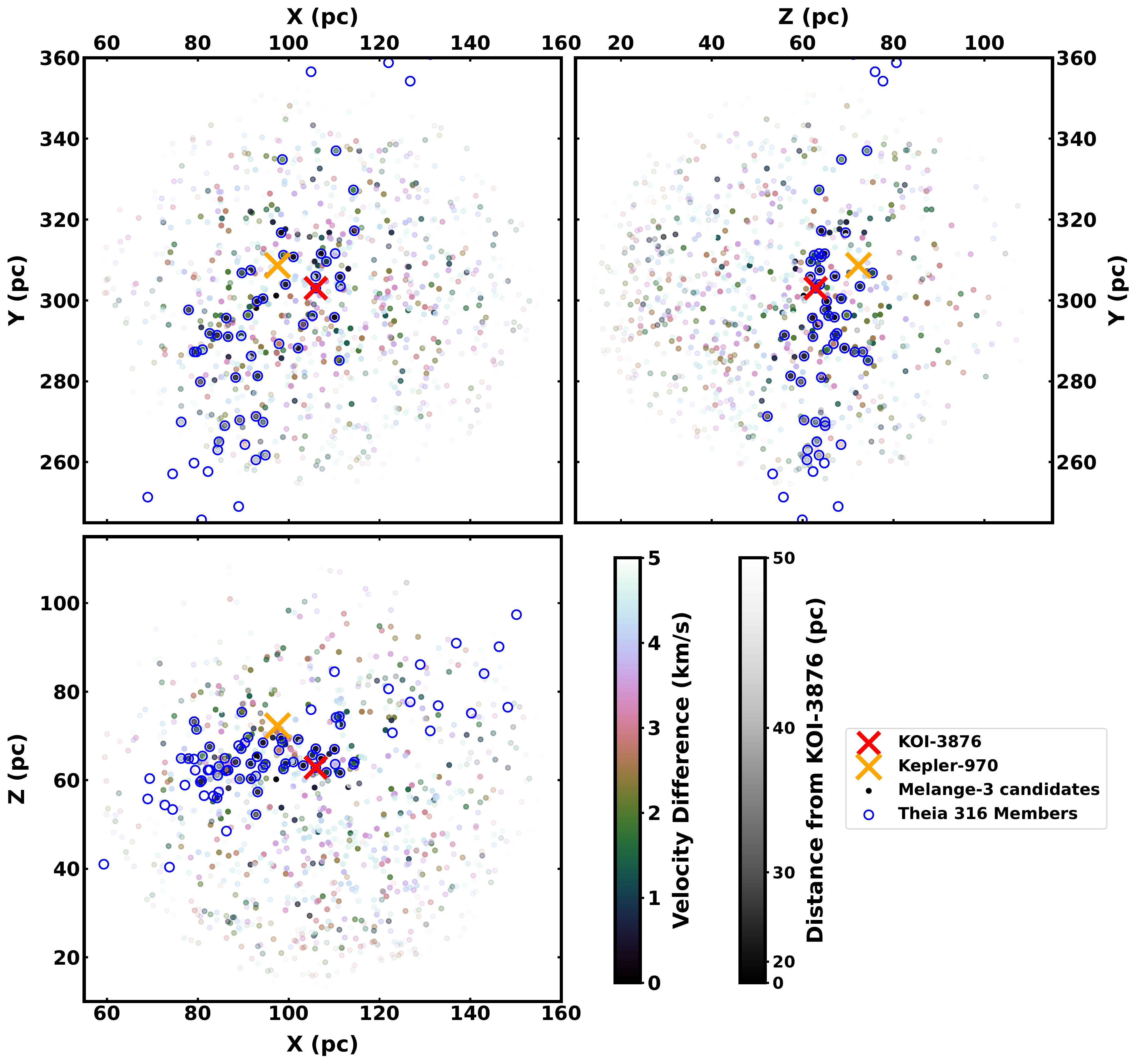}
    \caption{Galactic Heliocentric (XYZ) coordinates of stars within $D < 50$\,pc and $\Delta v_{tan} < 5$\,\kms\ of \starname, color-coded by their velocity difference from \starname\ and transparency set by their 3D separation from \starname. \starname\ is shown as a red X, Kepler-970 as an orange X, and members of Theia 316 from \citet{2019AJ....158..122K} as blue hollow circles. There is a high overlap between Theia 316 members and the cluster of \association\ candidates co-moving with and near \starname. A majority of high-probability \association\ candidates are $\geq$10\,pc from the Theia 316 string in low densities that would likely not trigger the clustering method used in \citet{2019AJ....158..122K}.  }
    \label{fig:xyz}
\end{figure*}

For each of the 1007 candidate members of \association, we show the proper motion in Galactic coordinates ($l$, $b$) in Figure~\ref{fig:ProperMotion} and the Galactic $XYZ$ position in Figure~\ref{fig:xyz}. While there is a clear overdensity of sources within $\simeq$1\,\kms\ of \starname, our initial selection of stars within $D < 50$\,pc and $\Delta v_{tan} < 5$\,\kms\ of \starname\ included a large number of field interlopers. A tighter cut on velocity and distance would lower contamination, but some sources far in separation have matching velocities and show other signs of youth (such as rotation). This made it challenging to derive the $UVWXYZ$ parameters of the association without further cuts on the full membership list.

We selected a set of high probability members independent of their \gaia\ parameters that were used for initial selection and focus on membership based on rotation periods and radial velocities. We included only sources with a rotation period consistent with the Pleiades sequence (as described in Section~\ref{sec:contam}) and radial velocities within $\Delta v_{rad} < 8$ \kms\ of the source. We removed three additional targets that had low-precision velocities (from LAMOST). Unlike when fitting the isochrones, the low-mass stars offer no particular advantage. 

The cuts left us with 31 high-probability members. From these, we estimate the Galactic position ($XYZ$) and velocity ($UVW$) of the association. We also calculated the intrinsic scatter in each parameter ($\sigma_X$, $\sigma_Y$, $\sigma_Z$, $\sigma_U$, $\sigma_V$, and $\sigma_W$) after accounting for measurement errors. The resulting parameters for \association\ are given in Table~\ref{tab:clusterparams}.

\begin{deluxetable}{lcccc}
\centering
\tabletypesize{\scriptsize}
\tablewidth{0pt}
\tablecaption{Properties of \association \label{tab:clusterparams}}
\tablehead{\colhead{Parameter} & \colhead{Mean} & \colhead{$\sigma$} }
\startdata
\hline 
$X$ (pc)  & 103 & 19 \\
$Y$ (pc)  & 289 & 18 \\
$Z$ (pc) & 63 & 14 \\
$U$ (\kms) & -9.2 & 1.4 \\
$V$ (\kms) & -25.54 & 0.36 \\
$W$ (\kms) & 1.0 & 1.2 \\
\enddata
\end{deluxetable}

A more formal treatment would account for covariance between parameters and derive the full two-dimensional matrix \citep[e.g.,][]{BanyanSigma}. However, the group may be part of a much larger stellar string, which would require a different model \citep{2019AJ....158..122K, 2021ApJ...915L..29G}. We discuss this possibility in more detail in Section~\ref{sec:theia}. Since we were primarily interested in characterizing the planets, we defer a more detailed analysis of \association\ and its relation to other groups for a future analysis focused on the association.

\subsection{Contamination Rate}\label{sec:contam}

As discussed above, we expect many of our 1007 candidate members are field interlopers. We were able to make a more quantitative estimate of the contamination rate using our rotation and radial velocity measurements. These were particularly effective because neither metric was used in our initial sample selection (\texttt{FriendFinder} only used the velocity of \starname).

We first estimated the contamination rate from the ratio of the number of stars with rotation periods consistent with the Pleiades sequence compared to the total number of stars for which we could estimate a reliable rotation period.

For the numerator, we used the Pleiades-like rotation cuts explained in Section~\ref{sec:rotation}. This yielded 92 stars we considered consistent with the Pleiades rotation sequence. The denominator is effectively the number of stars that would pass the above cut if every single star was a member. We assumed we could extract a rotation period for any member with a \kepler\ light curve. For \tess, we assumed we are complete out to $T_{\rm{mag}}<14$. This method suggested that 35\% of the candidates are true members. However, both assumptions were optimistic, particularly for \tess. Recovery rates for the Pleiades are closer to 90\% \citep{Rebull2016}; that survey used higher-quality \ktwo\ data (longer baseline and larger telescope) and Pleiades is twice as close to the Sun as the stars considered here. A more reasonable completeness of $\simeq$70\% would bring the true-member rate to 50\%.

A more accurate test can be done using radial velocities by testing what fraction of the stars are consistent with the association velocity in Table~\ref{tab:clusterparams}. Assuming an internal velocity dispersion of 0.5--1\kms\ (Table~\ref{tab:clusterparams}), 52\%--57\% of the candidate members with velocities are within 3$\sigma$ of the expected radial velocity for membership. As with rotation periods, some number of these will overlap with the distribution by chance. But the semi-uniform distribution of velocities outside the main locus seen in Figure~\ref{fig:groupRV} suggests this number is $<5$\%. We may miss a similar number of real members due to RV variation from binarity. 

The two estimates are broadly consistent with each other given the uncertainties and complicating factors. We conclude $\gtrsim$50\% of the 1007 candidates in our list are real members of \association. 

\subsection{Connection to Theia 316}\label{sec:theia}

\citet{2019AJ....158..122K} identified several young stellar associations from \gaia\ DR2 data. One such string, Theia 316, includes \starname. \citet{2019AJ....158..122K}, using a convolutional neural network in combination with traditional isochronal fittings, estimate an age for Theia 316 of 108\,Myr. The overlap in members and similar age estimates suggested that \association\ is a component of the larger Theia 316 string. 

As we show in Figure \ref{fig:xyz}, the populations show significant overlap in $XYZ$ space. Many stars in Theia 316 were missing from our list of \association\ candidates, the great majority of which were more than 50\,pc from \starname\ (our \texttt{FriendFinder} adopted search radius). The density clustering method used in \citet{2019AJ....158..122K} had no restrictions on the physical size of the group, as long as the number of sources in a given spatial and proper motion space is sufficiently dense. By contrast, \texttt{FriendFinder} was designed to help age-date a specific star, and hence identify targets within a given radius in kinematic and physical space. Increasing the search radius on \texttt{FriendFinder} would likely help us recover some of the missing stars, but also increase the contamination rate, likely not improving our ability to assign ages to the two planet hosts. 

Similarly, many high-probability members of \association\ were missing from the Theia 316 list (including \starnametwo). Forty stars match expectations for membership in \association\ for at least two of the following: rotation period, radial velocity, and lithium levels. These 40 should have had a relatively low field contamination rate. However, only 15 of the 40 were on the Theia 316 list. Adjustments to this high-probability list (e.g., tighter cuts on radial velocity, requiring all three of the criteria to match) yielded almost identical conclusions: slightly more than half the real members were missing from the Theia 316 list. Most of these missing stars landed far from the core of the string in $XYZ$ space and likely failed to pass the minimum density requirements. Even for our analysis, we would not have called most of these clear members without the additional rotation or spectroscopic data.

\section{Search for planets in \association}\label{sec:search}

To check for other candidate planets or eclipsing binaries in the same association, we searched \kepler\ and \tess\ light curves using the Notch pipeline\footnote{\url{https://github.com/arizzuto/Notch_and_LOCoR}} as described in \citet{Rizzuto2017}. To briefly summarize, the Notch filter fits a window of the light curve as a combination of an outlier-robust second-order polynomial (for the stellar variability) and a trapezoidal notch (representing the potential planet). The window moved along the light curve until the variability is detrended (flattened) while preserving the planet signal. At each data point, we calculated the improvement from adding the trapezoidal notch based on the change in the Bayesian Information Criterion (BIC) compared to modeling just a polynomial. Notch then repeated this process over the full curve. 

We searched the detrended light curves and the BIC signals that Notch produces for periodic signals. We excluded signals at the stellar rotation period and its harmonics, which arise from imperfect detrending. We also checked for known \kepler\ Objects of Interest (KOIs) orbiting candidate members of \association, identified by a simple cross-match against the most recent KOI catalog \citep{2016AJ....152..158T}. All but one KOI were flagged by the Notch search. 

In total, we identified 19 targets of interest that pass the SNR and initial quality checks from Notch. Nine of these are known KOIs (KOI-678.01, KOI-678.02, KOI-966.01, KOI-966.02, KOI-1838.01, KOI-3876.01, KOI-5304.01, KOI-6819.01, and KOI-7059.01), while the remaining are newly identified. All 19 targets are listed in Table~\ref{tab:kois-kics} along with our classification of each. 

\subsection{Discussion of Individual Candidate Planet Hosts}

We split the candidate planet hosts into likely (Section~\ref{sec:acceptedMems}) and unlikely (Section~\ref{sec:rejectedMems}) members  of \association. For each system, we discuss our reasons for accepting or rejecting the candidate planet, based on \association\ membership and the quality of the planet signal. Other than \starname, we identified only one other system (\starnametwo) that was both likely to host a real planet and be a member of \association. 

\begin{deluxetable}{lcccrrr}
\centering
\tabletypesize{\scriptsize}
\tablecaption{Planetary Candidates in \association. \label{tab:kois-kics}}
\tablehead{\colhead{ID} & \colhead{Disposition} & \colhead{Mem?} & \colhead{$P_{\rm{orb}}$} & \colhead{$T_0$} & \colhead{Depth }\\[-0.3cm]
\colhead{} & \colhead{} & \colhead{} & \colhead{(days)} & \colhead{(MBJD)} & \colhead{(ppm)}
}
\startdata
\hline
KOI-678.01 & Conf & N & 6.040 & 55005.094 & 241\\ 
KOI-678.02 & Conf & N & 4.139 & 54964.516 & 256\\
KOI-966.01 & FP & Y & 0.379 & 55002.748 & 1973\\
KOI-966.02 & FP & Y & 7.769 & 54965.590 & 1730\\
KOI-1838.01 & Conf & Y & 16.736 & 54975.632 & 1147\\
KOI-3876.01 & Conf & Y & 19.578 & 54964.215 & 450\\
KOI-5304.01 & FP & N & 206.310 & 55064.318 & 277\\
KOI-6819.01 & Cand & N & 3.226 & 54966.919 & 36\\
KOI-7059.01 & EB & N & 31.973 & 54966.367 & 75053\\
\hline
KIC~6366739 & Cand & N & 22.321 & 55386.107 & 2277\\
TIC~20352534 & FP & N & 6.377 & 58687.371 & 5852\\ 
TIC~237181417 & FP & N & 10.870 & 58690.062 & 3178 \\ 
TIC~272486188 & FP & N & 9.040 & 58683.104 & 779\\ 
TIC~28768382 & FP & N & 11.790 & 58692.223 & 1430\\ 
TIC~138966713 & FP & Y & 19.701 & 58687.157 & 1858\\ 
TIC~158168145 & FP & Y & 7.107 & 58689.209 & 1908\\ 
TIC~158415341 & FP & N & 20.016 & 58690.003 & 7571\\ 
TIC~164678171 & FP & N & 12.392 & 58695.052 & 1277\\ 
TIC~355909811 & FP & N & 6.065 & 58686.431 & 10753\\ 
\enddata
\tablecomments{$T_0$ and $P$ for KOIs taken from ExoFOP-\kepler\ \citep{EXOFOP_Kepler}. }
\end{deluxetable}

\subsubsection{Candidate planets accepted as \association\ Members}\label{sec:acceptedMems}

\planetname\ was the original seed around which we identified \association, but it was still useful to confirm it is a member of the group. With a rotation period of 4.67\,days, \starname\ was an excellent match to the Pleiades sequence for its color. The EW[Li] ($134\pm18$\AA) was similarly consistent with the Pleiades distribution. Re-running \texttt{FriendFinder} on other high-probability members of \association\ also returned \starname\ in the membership list, confirming that kinematic and spatial agreement was robust to the initial seed. Lastly, the star is listed in the Theia 316 membership list, affirming its membership. We confirmed the signal is planetary in origin in Section~\ref{sec:fpp}. 

KOI-1838.01 is a confirmed planet \citep[Kepler-970 b;][]{2016ApJ...822...86M}. The star's rotation (9.23\,days) places it right on the Pleiades sequence (see Figure~\ref{fig:rotation}), and much faster than the stalling regime seen for $>600$\,Myr systems \citep{Curtis_stall}. The LAMOST (corrected) velocity is -32$\pm$4\,\kms, which is consistent with the value predicted for membership ($\simeq$-27\kms). A more precise measurement from the CKS-cool project \citet{CKS_cool} yielded an RV of -27.15$\pm$0.10\kms, an excellent match for the association. The corresponding high-resolution spectrum from \citet{CKS_cool} shows weak lithium ($<40$m\AA), but this is consistent with 100\,Myr for its spectral type \citep{BHAC15}. As we show in Figures~\ref{fig:CMD}, \ref{fig:ProperMotion}, and \ref{fig:xyz}, \starnametwo\ lands on the expected CMD for a Pleiades-aged star and is nearby other likely members both in Galactic position and proper motion. We classified this planet as real and a member; we included it in our analysis throughout the rest of this paper. 

Both planet candidates around KOI-966 are flagged as false positives by the \kepler\ analysis \citep{2016ApJS..224...12C, EXOFOP_Kepler}, with both signals attributed to an eclipsing binary (.02 corresponding to a harmonic of the secondary eclipse). Initially, we came to the same conclusion based on the V-shaped signal and a visible secondary eclipse. However, the depth is unusually small for an eclipsing binary ($\simeq$2\,mmag), the depth changes from quarter to quarter, and no matching signal is present in the \tess\ data (it should be marginally detectable at this magnitude). It is possible that this is a more exotic object with an evolving transit depth, but more likely is that this is due to a column anomaly or CCD cross-talk from another eclipsing binary somewhere else on the \kepler\ field \citep[for more details see][]{2014AJ....147..119C}. We recovered the same rotation period in the \tess\ and \kepler\ data (3.9\,days), which matches expectations for membership. The star is also only $\Delta v_{tan} \simeq 1$\kms\ and $D \simeq 8$\,pc from \starname, near the core of likely members. We classified this system as a member of \association but the signal as a false positive.

Using Notch, TICs 138966713 and 158168145 passed the BIC criterion for significance, but upon further visual inspections, we classified these as false positives. The transit signals did not have the expected transit shape and could be more easily explained as systematics or imperfectly removed stellar noise. TIC 138966713 has a radial velocity of $-29.9 \pm 10.0$ \kms, consistent with the expected radial velocity. TIC 158168145 has a radial velocity $-24.4 \pm 1.1$ and a rotation period of $5.4 \pm 0.5$, both of which are consistent with the expected values. We classified both targets as members. 

\subsubsection{Candidate planets rejected as \association\ Members}\label{sec:rejectedMems}

KOI-678 (.01 and .02) contains two confirmed planets (Kepler-211\,bc), and the star's light curve showed a clear rotation signature of $\simeq$13.7\,days. However, for a member of the same color, we expected the rotation period to be $\lesssim$8\,days. While the detected period could have been a harmonic, the EW[Li] measurement was only 3.8m\AA\ \citep{2018ApJ...855..115B}, while membership predicted a Li level above 100m\AA. The star's proper motion also put it on the outskirts of the distribution. We classified this star as a non-member. 

KOI-5304.01 was flagged as a false positive by the \kepler\ analysis due to a non-transit-like shape. Our visual inspection of the candidate agrees with this. The host star also shows a measurable but slow rotation period of $\simeq$11.4 days, which is slower than expected for membership (see Section~\ref{sec:rotation}). We classified this system as a non-member and the signal as a false positive.

KOI-6819 (.01) contains a single planet signal that passed visual inspection. However, the star showed no significant rotation, and the \gaia\ DR2 velocity was $\simeq$20\kms\ from the expected value for membership in the association. We classified it as a non-member and did not attempt to further validate the transit signal.

KOI-7059.01 was flagged as a false positive by the \kepler\ analysis \citep{2016ApJS..224...12C}. \citet{2014MNRAS.443.3068B} characterize this system as an eclipsing binary with both ellipsoidal variations and pulsations. They found the system age to be 900$\pm$200\,Myr, consistent with membership. However, the orbital solution from \citet{2014MNRAS.443.3068B} gave a systemic radial velocity $>20$\kms\ from the expected value for membership. The source also lands more than 40\,pc from the core of the group. So we decided to classify this as a non-member eclipsing binary. 

\begin{figure}[tb]
    \centering
    \includegraphics[width=0.49\textwidth]{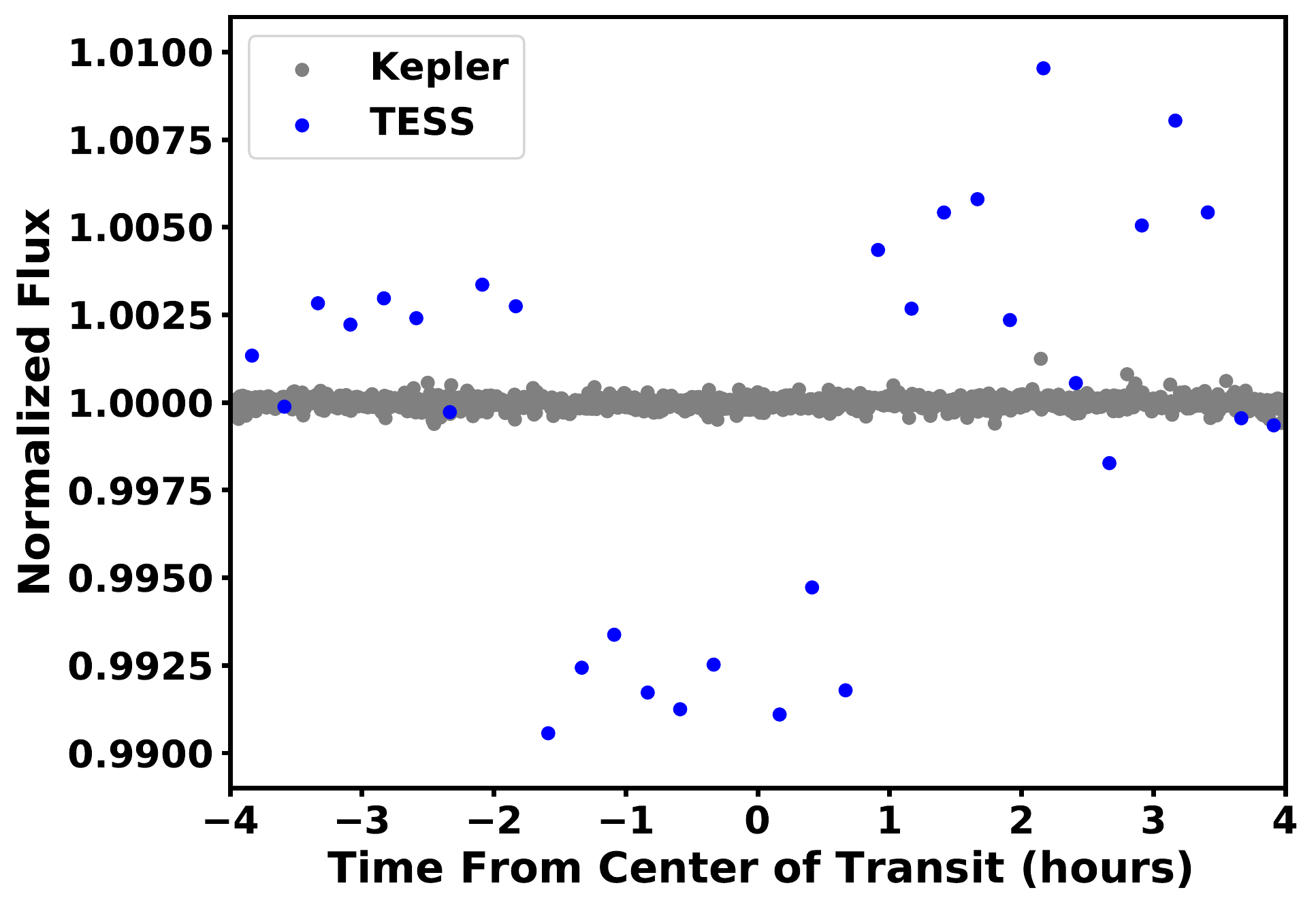}
    \caption{Phase-folded light curve of potential member TIC 158415341 (KIC 9700914) from \tess\ (blue) and \kepler\ (grey) detrended with the Notch filter. We found TIC 158415341 has a best fit period of 20.016 days with an initial transit time of 1690.503 BTJD using \tess\ data, but we fail to detect the transit using \kepler\ data. We reject this signal as noise in the \tess\ data.}
    \label{fig:interestingtransits}
\end{figure}

The newly identified targets, KIC 6366739 and TICs 20352534, 237181417, 272486188, 28768382, 158415341, 164678171, and 355909811, were all initially flagged by Notch as potentially interesting signals. TIC 272486188, 164678171, and 158415341 have both \tess\ and \kepler\ data, but we could not recover the transit signals we identified from \tess\ data in the higher-quality \kepler\ data. Figure \ref{fig:interestingtransits} shows TIC 158415341 as an example of this; there was a potentially interesting transit in \tess, but we failed to recover TIC 158415341's (KIC 9700914's) transit signal in the \kepler\ data. We rejected these targets as false positives. KIC 6366739 shows signs of a possible signal in the \kepler\ photometry. However, the target has a long period (22\,days) and was observed for only a single quarter, so the SNR was insufficient to reject or confirm this. When using Notch, TICs 20352534, 237181417, 28768382, and 355909811 passed the BIC criterion for significance, but upon further visual inspections, we find the transit signals did not have the expected transit shape and are more easily explained as systematics or imperfectly removed stellar noise. We classified these as FPs.

TICs 272486188, 158415341, and 164678171 are unlikely to be members of \association\ as the radial velocity measurements are $>$20\kms\ from the predicted values for membership. Additionally, KIC 6366739 is unlikely to be a member since the rotation period is $16.9 \pm 1.3$ days, which is greater than the expected rotation periods for this association ($<10$\,days). 

TICs 20352534, 237181417, and 355909811 did not have available rotation periods, radial velocities, or Lithium measurements, but we concluded all are unlikely to be members. While TIC 20352534 is likely young since it is an A-type, the star landed on the edge of the search region (49.34 pc from \starname) and the edge of the tangential velocity difference allowance (4.45 \kms\ from \starname). TICs 237181417 and 355909811 showed little flux variation, and therefore no signs of rotation, indicating they are most likely older stars not a part of the association.

\section{Properties of \starname\ and \starnametwo }\label{sec:stellar_params}

We summarize the properties of both host stars in Table~\ref{tab:prop}, the details of which we provide below.

\subsection{Literature Parameters}
As a reasonably bright ($K_P=12.6$) star hosting a planet candidate from the \kepler\ mission, \starname\ has numerous stellar parameters in the literature. The California \kepler\ Survey (CKS) estimate \teff\ = $5720\pm60$\,K, \logg = 4.64$\pm$0.1, and \vsini\ = 9.9$\pm$1.0\kms, $R_*=0.95^{+0.06}_{-0.04}R_\odot$ and $M_*=1.01\pm0.03M_\odot$ based on comparing their high-resolution spectra and comparison to well-characterized templates \citep{2017ApJ...836...77Y, 2017AJ....154..107P} and stellar isochrones \citep{JohsonCKS2017}. \citet{2018ApJS..237...38B}, using the same spectra, estimate \teff\ = 5642$\pm$27\,K, \logg = $4.46\pm$0.05, $R_*=0.93\pm0.02M_\odot$, $M_*=0.99\pm0.02M_\odot$, and \vsini\ = 10.4$\pm$0.5\kms, as well as detailed abundances that are generally consistent with the Solar value. \citet{2020AJ....159..280B} incorporated \gaia\ DR2 data with MIST stellar isochrones to derive an \teff\ = $5577\pm85$\,K, \logg $=4.50\pm0.02$, and $R_*=0.908\pm0.017R_\odot$. 

For \starnametwo, \citet{2020AJ....159..280B} estimated \teff\ = 4314$\pm$73\,K, \logg$=4.63\pm0.02$, $R_*=0.649\pm0.012R_\odot$, and $M_*=0.657\pm0.022M_\odot$ using the \gaia\ parallax, photometry, and MIST isochrones as with \starname. Based on the CKS spectra, \citet{CKS_cool} found a consistent \teff\ (4401$\pm$70\,K), stellar radius ($R_*=0.70\pm0.03R_\odot$), and mass ($M_*=0.68\pm0.10M_\odot$). \citet{CKS_cool} also estimated a \vsini\ of 5.21$\pm$1.0\kms\ from the CKS spectra. This \vsini\ may be slightly overestimated due to mild broadening in their templates, but the errors are sufficiently generous for our analysis. 

These stellar parameters are generally in agreement with each other. However, literature estimates that relied on model isochrones \citep[e.g.,][]{2020AJ....159..280B} assigned $>2$\,Gyr ages for both stars, much older than the true $\simeq$100\,Myr age from the association. Although this will not impact purely spectroscopic parameters like \teff\ and \vsini, the assumption can have a strong impact on the estimated stellar mass. Thus, we revisit these parameters with our analysis below.

\subsection{Spectral-Energy Distribution}\label{sec:sed}

\begin{figure}[tb]
    \centering
    \includegraphics[width=0.49\textwidth]{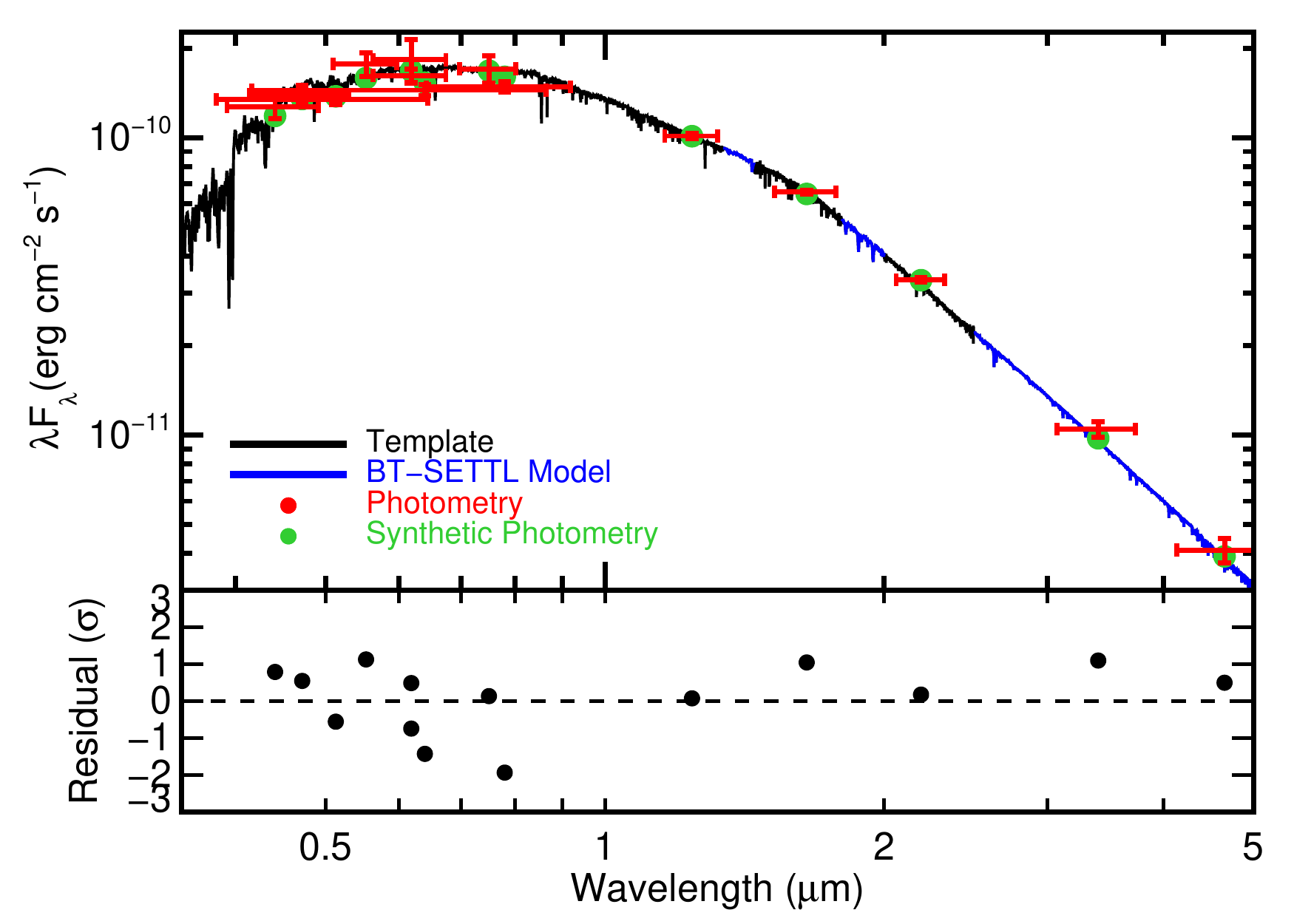}
    \caption{Best-fit template spectrum (G1V; black) and synthetic photometry (green) compared to the observed photometry of \starname\ (red). Errors on observed photometry are shown as vertical error bars, while horizontal error bars indicate the approximate width of the filter. BT-SETTL models (blue) were used to fill in regions of high telluric absorption or beyond the template range. The bottom panel shows the photometric residual in units of standard deviations. }
    \label{fig:sed}
\end{figure}

We fit the observed spectral-energy-distribution (SED) of both stars following \citet{Mann2016b}. To briefly summarize, we fit the observed photometry with a grid of optical and near-infrared flux-calibrated spectra spanning 0.4--2.3\um. We included BT-SETTL CIFIST atmospheric models \citep{BHAC15} in the fit, both to estimate the \teff\ and fill in gaps in the template spectra (e.g., beyond 2.3\um). We integrated the resulting absolutely-calibrated spectrum to estimate the bolometric flux (\fbol), which we combined with the \gaia\ EDR3 parallax to estimate the stellar luminosity ($L_*$). With \teff\ and $L_*$, we calculated $R_*$ using the Stefan-Boltzmann relation. While reddening in this sight-line is low \citep{2011ApJ...737..103S}, \starname\ is well outside the Local Bubble, so we included extinction as part of the fit. To account for variability in the star, we added (in quadrature) 0.02 mags to the errors of all-optical photometry. In total, the fit included six free parameters: the spectral template, $A_V$, three parameters that describe the model ($\log~g$, \teff, and [M/H]), and a scale factor between the model and the photometry. We show an example fit in Figure~\ref{fig:sed}. 

For \starname, the resulting fit yielded $A_V=0.16^{+0.10}_{-0.08}$, \teff=$5672\pm65$\,K, \fbol=$(2.55\pm0.10)\times10^{-10}$ (erg\,cm$^{-2}$\,s$^{-1}$), $L_*=0.81\pm0.03L_\odot$, and $R_*=0.94\pm0.03R_\odot$. For \starnametwo, we found $A_V=0.28\pm0.12$, \teff=$4290\pm70$\,K, \fbol=$(4.88\pm0.23)\times10^{-11}$ (erg\,cm$^{-2}$\,s$^{-1}$), $L_*=0.168\pm0.010L_\odot$, and $R_*=0.715\pm0.03R_\odot$. 

Our SED parameters were in good agreement with the literature spectroscopic values for both stars. Since \starname\ is Sun-like, we considered the CKS spectroscopic \teff\ to be more reliable than the SED-based value, but the SED-based luminosity (and radius) is more reliable than the one derived from the spectroscopic \logg\ or isochrone. We combined the two, which yielded a final radius of $0.92\pm0.02R_\odot$. For \starnametwo, we adopted our SED-fit parameters for all values. 

\citet{Tayar2022} argue that current interferometric diameter data suggests a limit on the precision of 2\% on $L_*$ and 4\% on $R_*$ for Sun-like stars. Our radius for Kepler-970 is consistent with this limit, but our estimated radius for \starname\ had an error of only 2\%. We conservatively adopted the larger 4\% uncertainty ($0.04R_\odot$). 

\subsection{Stellar mass}\label{sec:mass}

To determine $M_*$ for \starname, we compared the observed photometry to Solar-metallicity magnetic DSEP evolution models and PARSEC models. We used \texttt{emcee} to simultaneously fit for age, $A_V$, $M_*$, and an additional parameter to capture underestimated uncertainties in the data or models ($f$, in magnitudes) within an MCMC framework. We used a hybrid interpolation method, first identifying the nearest age in the model grid and then performing a linear interpolation in mass to obtain stellar parameters and model photometry. Since this method could not interpolate between ages, we re-sampled the input grid using the \texttt{isochrones} package \citep{2015ascl.soft03010M} to be more dense (0.1\,Myr and 0.01$M_\odot$) than expected errors. To redden the model photometry, we used \texttt{synphot} \citep{pey_lian_lim_2020_3971036} and the extinction law from \citet{1989ApJ...345..245C}. We placed a Gaussian prior on the age of 105$\pm$10\,Myr, while other parameters evolved under uniform priors. The resulting fit from each model grid was very precise, but differences between the two grids suggest larger systematic errors. Considering these, the resulting parameters were generally in agreement with our spectroscopic constraints ($R_*=0.968\pm0.07$, $A_V=0.27\pm0.10$, \teff=$5710\pm60$) and provided a stellar-mass estimate of $M_*=1.04\pm0.03M_\odot$. We combined this with our earlier radius estimate to get an estimate of the stellar density ($\rho_*=1.30\pm0.18\rho_\odot$).

\starnametwo\ is expected to be on the main-sequence (see Figure~\ref{fig:CMD}) and is low-mass enough to fit within the bounds of the $M_K-M_*$ relation from \citet{Mann2019a}. We used the \gaia\ EDR3 parallax with uncertainties corrected for underestimated errors following \citet{2021MNRAS.506.2269E} and the 2MASS $K_S$ magnitude. We applied a correction for reddening based on our SED fit (Section~\ref{sec:sed}), although this was negligible in the $K_S$ band compared to other uncertainties. We fed the resulting values into the fit posteriors using the provided code\footnote{\url{https://github.com/awmann/M_-M_K-}}, which yielded a mass of $0.66\pm0.02M_\odot$. Since the star is near the zero-age main-sequence, we checked our mass estimate using the same methodology as with \starname. The model-based mass was consistent but showed more variation between grids ($0.65-0.70M_\odot$), so we adopted the value from the M$_K-M_*$ relation. Combining with the radius above, this gives a stellar density of $1.88\pm0.24\rho_\odot$.

\begin{deluxetable*}{lcccc}
\centering
\tabletypesize{\scriptsize}
\tablewidth{0pt}
\tablecaption{Properties of \starname\ and \starnametwo\ \label{tab:prop}}
\tablehead{\colhead{Parameter} & \colhead{\starname} & \colhead{\starnametwo} & \colhead{Source} }
\startdata
\multicolumn{4}{c}{Name}\\
\hline 
Gaia EDR3 & 2052827207364859264 & 2101379205604338688 & \citet{GaiaEDR3}\\
KOI & 3876 & 1838 & \citet{2016AJ....152..158T} \\
KIC & 3440118 & 5526527 & \citet{Brown2011} \\
TIC & 122450696  & 122069243 & \citet{TIC2018} \\
2MASS & J19214575+3831248 & J19183005+4042314 & \citet{Skrutskie2006} \\
\hline
\multicolumn{4}{c}{Astrometry}\\
\hline
 $\alpha$  &  290.440629 & 289.625218 & \emph{Gaia} EDR3\\
 $\delta$. & 38.523572  & 40.708735 & \emph{Gaia} EDR3 \\ 
 $\mu_\alpha$ (mas\,yr$^{-1}$)& -4.154$\pm$0.010 & -2.739$\pm$0.020 & \emph{Gaia} EDR3\\ 
 $\mu_\delta$  (mas\,yr$^{-1}$) & 2.269$\pm$0.011 & 1.652$\pm$.024 & \emph{Gaia} EDR3\\ 
 $\pi$ (mas) & 3.0565$\pm$0.0093 &3.0153$\pm$0.0186 & \emph{Gaia} EDR3\\ 
\hline
\multicolumn{4}{c}{Photometry}\\
\hline
$G_{Gaia}$ (mag) & 12.6054$\pm0.0028$ &14.7876$\pm$0.0029 & \emph{Gaia} EDR3\\
$BP_{Gaia}$ (mag) & 12.9642$\pm0.0033$ & 15.4884$\pm$0.0049 & \emph{Gaia} EDR3\\ 
$RP_{Gaia}$ (mag) & 12.0798$\pm0.0041$ & 13.9759$\pm$0.0048 & \emph{Gaia} EDR3\\
$B$ (mag) & $13.375 \pm 0.094$ & \ldots & APASS \\ 
$V$ (mag) & $12.655 \pm 0.122$ & \ldots & APASS \\ 
$g$' (mag) & $13.038 \pm 0.033$ & \ldots & APASS \\ 
$r$' (mag) & $12.456 \pm 0.092$ & \ldots & APASS \\ 
$i$' (mag) & $12.323 \pm 0.062$ & \ldots & APASS \\ 
$J$ (mag) & $11.456 \pm 0.02$ & 12.980$\pm$0.023 &  2MASS\\ 
$H$ (mag) & $11.152 \pm 0.016$  & 12.355$\pm$0.023 &2MASS\\	 
$K_S$ (mag) & $11.107\pm 0.019$ & 12.215$\pm$0.022 & 2MASS\\ 
$W1$ (mag) & $11.06 \pm 0.023$ & 12.155$\pm$0.022 & ALLWISE\\
$W2$ (mag)& $11.09 \pm 0.020 $ & 12.193$\pm$0.022 & ALLWISE\\
$W3$ (mag)& $10.91 \pm 0.094$ & 12.195$\pm$0.282 & ALLWISE\\
\hline
\multicolumn{4}{c}{Kinematics \& Position}\\
\hline
RV$_{\rm{Bary}}$ (km\,s$^{-1}$) & -26.79$\pm$0.01 & -27.15$\pm$0.10 & \citet{2020AJ....160..120J}; \citet{CKS_cool}\\
U (km\,s$^{-1}$) & -9.467 $\pm$ 0.016 & -8.815 $\pm$ 0.047 & This work\\
V (km\,s$^{-1}$) & -26.034 $\pm$ 0.012 & -26.146 $\pm$ 0.092 & This work\\
W (km\,s$^{-1}$) & $2.048 \pm 0.032$ & -1.034 $\pm$ 0.055 & This work\\
X (pc) & $105.98 \pm 0.42$ & $97.55 \pm 0.78$ & This work\\
Y (pc) & $303.03 \pm 1.20$ & $308.66 \pm 2.45$ & This work\\
Z (pc) & $62.86 \pm 0.25$ & $72.28 \pm 0.58$ & This work\\
\hline
\multicolumn{4}{c}{Physical Properties}\\
\hline
$P_{\rm{rot}}$ (days) &  $4.69\pm0.04$ & 9.23 $\pm$0.66& \citet{Nielsen:2013}; \citet{Santos2019}\\
\vsini (km\,s$^{-1}$) & $ 10.4\pm0.5 $ & 5.2$\pm$1.0\kms & \citet{2018ApJS..237...38B}; \citet{CKS_cool} \\
$i_*$ ($^\circ$) & $ >80$ & $>52$& This work\\
\fbol\,(erg\,cm$^{-2}$\,s$^{-1}$)& $(2.55\pm0.10)\times10^{-10}$ &$(4.88\pm0.23)\times10^{-11}$ & This work\\
\teff\ (K) & $5720 \pm 60$ & $4290\pm70$ & CKS; This work \\
\textup{[Fe/H]} & $0.12\pm0.02$ & 0.04$\pm$0.09 & \citet{2018ApJS..237...38B}; \citet{CKS_cool}\\
M$_\star$ (M$_\odot$) & $ 1.01\pm0.03 $ & 0.67$\pm$0.02 & This work \\
R$_\star$ (R$_\odot$) &  $ 0.92\pm0.04 $ & $0.71\pm0.03$ & This work \\
L$_\star$ (L$_\odot$) & $ 0.81\pm0.03 $ & $0.168\pm0.010$ & This work \\
$\rho_\star$ ($\rho_\odot$) & $1.30\pm0.18$ & 1.88$\pm$0.24 & This work \\
Age (Myr) & \multicolumn{2}{c}{$105\pm10$}   & This work
\enddata
\end{deluxetable*}

\subsection{Stellar inclination}\label{sec:inc}

To test whether the stellar spin and planetary orbit are consistent with alignment, we computed the stellar inclination ($i_*$) from the \vsini, $P_{\rm{rot}}$, and $R_*$ values estimated above. In its simplest form, this calculation is $V=2\pi R_*/P_{\rm{rot}}$, but requires additional statistical corrections \citep[see ][]{MortonWinn2014, 2020AJ....159...81M}. Here we followed the methodology described in \citet{2020AJ....159...81M}. For \starname, the resulting stellar inclination was $i_*>71^\circ$ at 95\% confidence and $i_*>80^\circ$ at 68\% confidence. For \starnametwo, the values were $i_*>52^\circ$ and $i_*>70^\circ$ at 95\% and 68\%, respectively. Both are consistent with alignment with the orbital inclinations. 

\section{Transit parameters}\label{sec:transit}

We fit the \kepler\ photometry using the \texttt{misttborn} (MCMC Interface for Synthesis of Transits, Tomography, Binaries, and Others of a Relevant Nature) fitting code\footnote{\url{https://github.com/captain-exoplanet/misttborn}} first described in \citet{Mann2016a} and expanded upon in \citet{MISTTBORN}. \texttt{misttborn} uses \texttt{BATMAN} \citep{Kreidberg2015} to generate model light curves and \texttt{emcee} \citep{Foreman-Mackey2013} to explore the transit parameter space. 

\begin{deluxetable*}{lccccc}
\centering
\tablewidth{0pt}
\tablecaption{Parameters of \starname\ and \starnametwo\ \label{tab:bestfitParams}
}
\tablehead{
\colhead{Parameter} & \multicolumn{2}{c}{\starname} & \multicolumn{2}{c}{Kepler-970}\\[-0.3cm] 
\colhead{} & \colhead{e=0} & \colhead{e float (preferred)} & \colhead{e=0} & \colhead{e float (preferred)}
}
\startdata
\multicolumn{5}{c}{Measured Parameters }\\
\hline
$T_0$ (BJD-2454833) & $131.71488 \pm 0.00088$ & $131.71489^{+0.00101}_{-0.00095}$ & $143.13258^{+0.00099}_{-0.00098}$ & $143.1326 \pm 0.0011$\\
$P$ (days) & $19.577831 \pm 2.1\times10^{-5}$ & $19.57783^{+2.2\times10^{-5}}_{-2.1\times10^{-5}}$ & $16.736525 \pm 2\times10^{-5}$ & $16.736525 \pm 2\times10^{-5}$ \\ 
$R_P/R_{\star}$ & $0.01945^{+0.00061}_{-0.00044}$ & $0.01984^{+0.00125}_{-0.00065}$ & $0.03186^{+0.00136}_{-0.00077}$ & $0.0337^{+0.0031}_{-0.0022}$\\ 
$b$ & $0.31^{+0.28}_{-0.22}$ & $0.55^{+0.25}_{-0.36}$ & $0.33^{+0.3}_{-0.23}$ & $0.7^{+0.18}_{-0.44}$\\ 
$\rho_{\star}$ ($\rho_{\odot}$) & $15.5^{+2.5}_{-5.9}$ & $1.3^{+0.19}_{-0.17}$ & $15.2^{+2.9}_{-6.7}$ & $1.882^{+0.09}_{-0.088}$ \\ 
$q_{1,1}$ & $0.289^{+0.107}_{-0.1}$ & $0.297^{+0.103}_{-0.097}$ & $0.37 \pm 0.12$ & $0.38 \pm 0.12$\\ 
$q_{2,1}$ & $0.371^{+0.075}_{-0.086}$ & $0.373^{+0.074}_{-0.088}$ & $0.374^{+0.07}_{-0.078}$ & $0.376^{+0.071}_{-0.076}$\\
$\sqrt{e}\sin\omega$ & -- & $0.39^{+0.18}_{-0.28}$  & -- & $0.25^{+0.25}_{-0.3}$\\ 
$\sqrt{e}\cos\omega$ & -- & $0.14^{+0.54}_{-0.69}$ & -- & $-0.0^{+0.55}_{-0.54}$\\ 
$\log(P_{GP})$ & $1.495^{+0.016}_{-0.014}$ & $1.494^{+0.016}_{-0.014}$ & $2.839^{+0.056}_{-0.053}$ & $2.838^{+0.056}_{-0.051}$\\ 
$\log(Amp)$ & $-9.338^{+0.057}_{-0.055}$ & $-9.338^{+0.056}_{-0.052}$ & $-6.797^{+0.079}_{-0.075}$ & $-6.799^{+0.078}_{-0.076}$\\ 
$\log(Q)$ & $1.154^{+0.058}_{-0.055}$ & $1.154^{+0.057}_{-0.055}$ & $0.664^{+0.029}_{-0.025}$ & $0.665^{+0.027}_{-0.025}$\\ 
\hline 
\multicolumn{5}{c}{Derived Parameters }\\
\hline
$a/R_{\star}$ & $76.2^{+3.9}_{-10.0}$ & $47.2^{+4.0}_{-6.9}$  & $68.2^{+4.0}_{-10.0}$ & $42.2^{+6.5}_{-7.6}$\\ 
$i$ ($^{\circ}$) & $89.76^{+0.17}_{-0.29}$ & $88.92^{+0.69}_{-0.4}$ & $89.72^{+0.2}_{-0.37}$ & $88.74^{+0.77}_{-0.26}$\\ 
$T_{14}$ (days) & $0.0793^{+0.0016}_{-0.0015}$ & $0.095^{+0.062}_{-0.032}$ & $0.0763^{+0.0021}_{-0.0018}$ & $0.09^{+0.047}_{-0.023}$\\ 
$T_{23}$ (days) & $0.0755^{+0.0015}_{-0.0016}$ & $0.087^{+0.06}_{-0.029}$ & $0.0703^{+0.0019}_{-0.0024}$  & $0.075^{+0.046}_{-0.022}$\\ 
$R_P$ ($R_\oplus$) & $1.953^{+0.075}_{-0.061}$ & $1.992^{+0.133}_{-0.078}$ & $2.467^{+0.148}_{-0.149}$ & $2.609^{+0.265}_{-0.203}$\\ 
$a$ (AU) & $0.313^{+0.046}_{-0.064}$ & $0.194^{+0.032}_{-0.039}$ & $0.225^{+0.016}_{-0.041}$ & $0.139^{+0.022}_{-0.026}$\\ 
$e$ & \ldots & $0.43^{+0.18}_{-0.14}$ & \ldots & $0.34^{+0.18}_{-0.22}$\\ 
$\omega$ ($^{\circ}$) & \ldots & $88.0^{+75.0}_{-59.0}$ & \ldots & $117.0^{+75.0}_{-85.0}$\\
\hline 
\enddata
\end{deluxetable*}

The standard implementation of \texttt{misttborn} fits for five parameters for each transiting planet: time of periastron ($T_0$), orbital period of the planet ($P$), planet-to-star radius ratio ($R_p/R_\star$), impact parameter ($b$), and stellar density ($\rho_\star$). We fit two linear and quadratic limb-darkening coefficients ($q_1$, $q_2$) following the triangular sampling prescription of \citet{Kipping2013}. 

We ran two versions of the fit. In the first, the MCMC chain restricted $e$ to 0 and allowed $\rho_\star$ to vary within a uniform distribution, and the second allowed $e$ to vary with a Gaussian prior on $\rho_\star$ from our spectroscopic and SED analysis (Section~\ref{sec:stellar_params}). For both fits, we applied Gaussian priors on the limb-darkening coefficients based on the values derived using our stellar parameters from Section~\ref{sec:stellar_params} and the \texttt{LDTK} toolkit \citep{2015MNRAS.453.3821P}, with errors accounting for the difference between models: $g_1=0.42\pm0.08$, $g_2=0.13\pm0.04$ for Kepler-1928 and $g_1=0.51\pm0.08$, $g_2=0.13\pm0.04$ for Kepler-970. Other transit parameters were sampled uniformly with physically motivated boundaries; e.g., $T_0$ was restricted to the time period sampled by the data, $|b|<1+R_P/R_*$, $0<e<1$, $0<\rho_*$, and $0<R_P/R_*<1$.

To model stellar variations, \texttt{misttborn} includes a Gaussian Process (GP) regression module, utilizing the \texttt{celerite} code \citep{celerite}. We initially used a mixture of two stochastically driven damped simple harmonic oscillators (SHOs) at periods $P_{GP}$ (primary) and $0.5P_{GP}$ (secondary), as is common for fitting stellar variability. However, we found that for both planets, the second SHO was not required (the model was weighting it to zero), and re-ran with a single SHO. There were three GP parameters: the log of the GP period ($\ln(P_{GP})$), the log of the GP amplitude ($\ln{\rm{Amp}}$), and a decay timescale for the variability (quality factor, $\ln{Q}$).  We used a weak Gaussian prior (20\%) on the GP period to keep walkers from wandering to half and double-period solutions. All other GP parameters evolved under uniform priors.

\begin{figure*}[tb]
    \centering
    \includegraphics[width=0.48\textwidth]{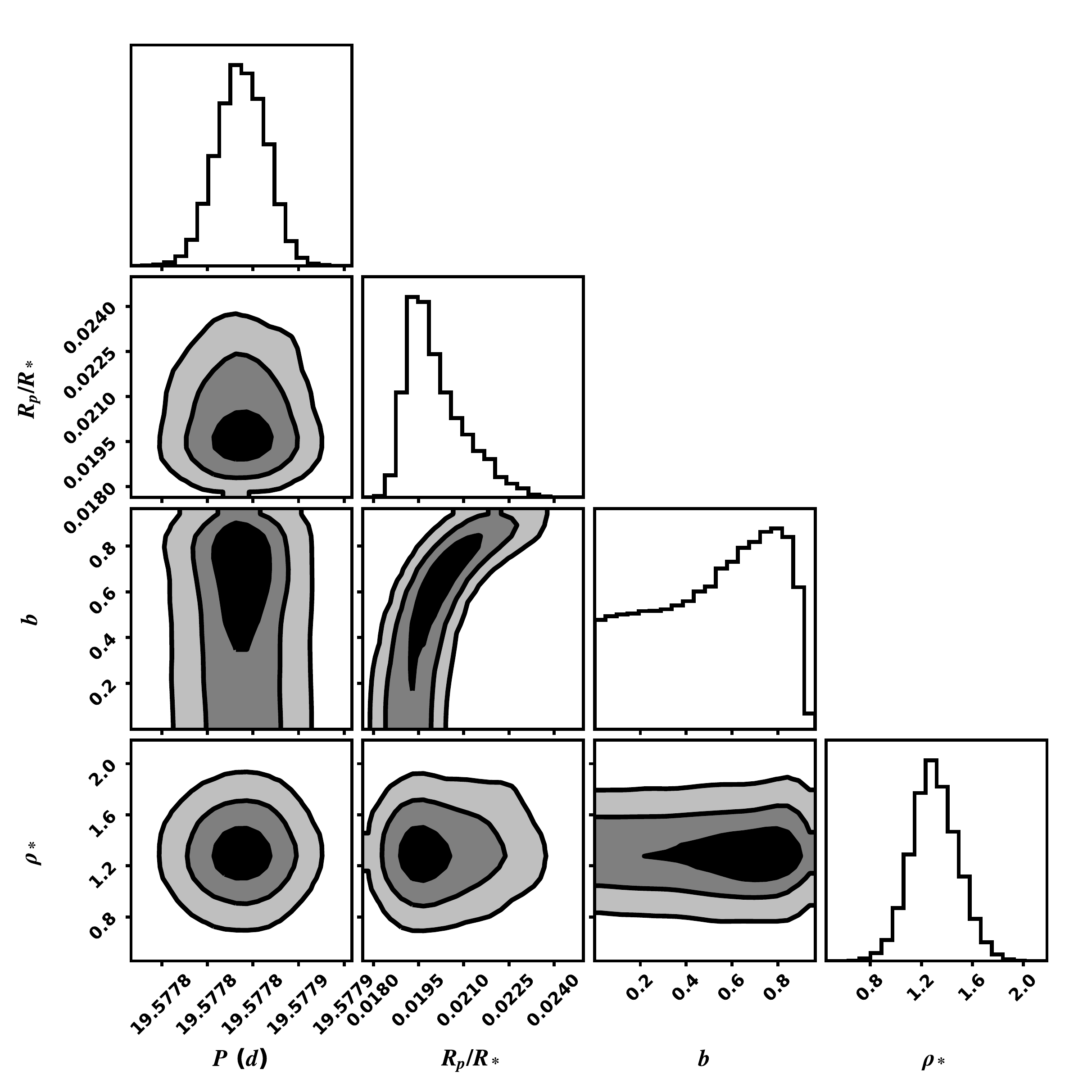}
    \includegraphics[width=0.48\textwidth]{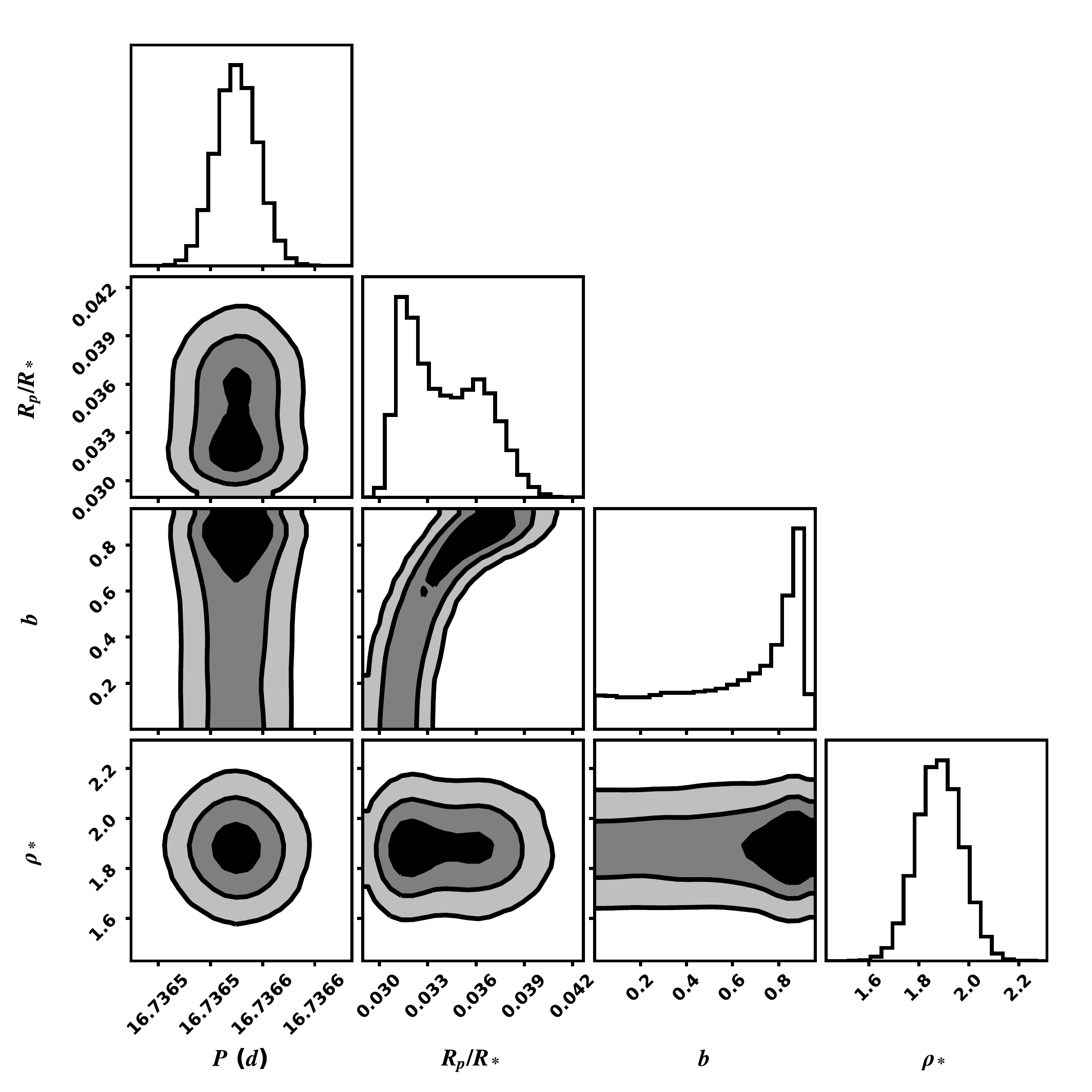}
    \caption{Corner plot of the major transit parameters ($P$, $R_P/R_*$, $b$, and $\rho_*$) from our \texttt{MISTTBORN} fit for \planetname\ (left) and \planetnametwo\ (right). The contour levels correspond to 1$\sigma$, 2$\sigma$, and 3$\sigma$ of the points (from darkest to lightest). The planet-to-star radius ratio and eccentricity are strongly covariant with impact parameter, as a higher impact parameter requires a deeper transit (and lower eccentricity) to reproduce the observed transit depth (and duration). It is difficult to break this degeneracy with \kepler\ long-cadence data alone, as the integration time is longer than the ingress/egress. Plot made using {\texttt corner.py} \citep{foreman2016corner}.}
    \label{fig:ecccorner}
\end{figure*} 

\begin{figure*}[tbh]
    \centering
    \includegraphics[width=0.48\textwidth]{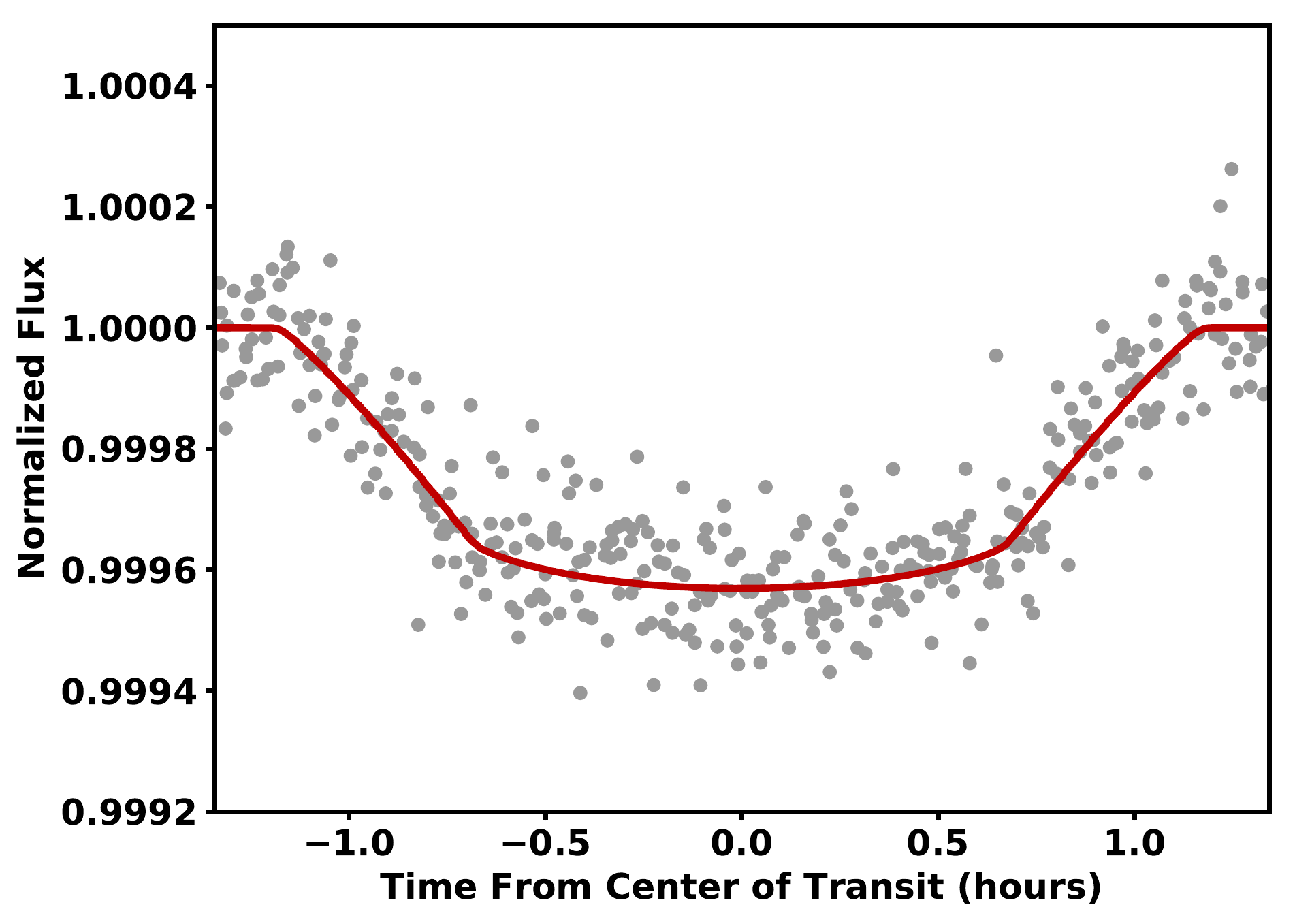}
    \includegraphics[width=0.48\textwidth]{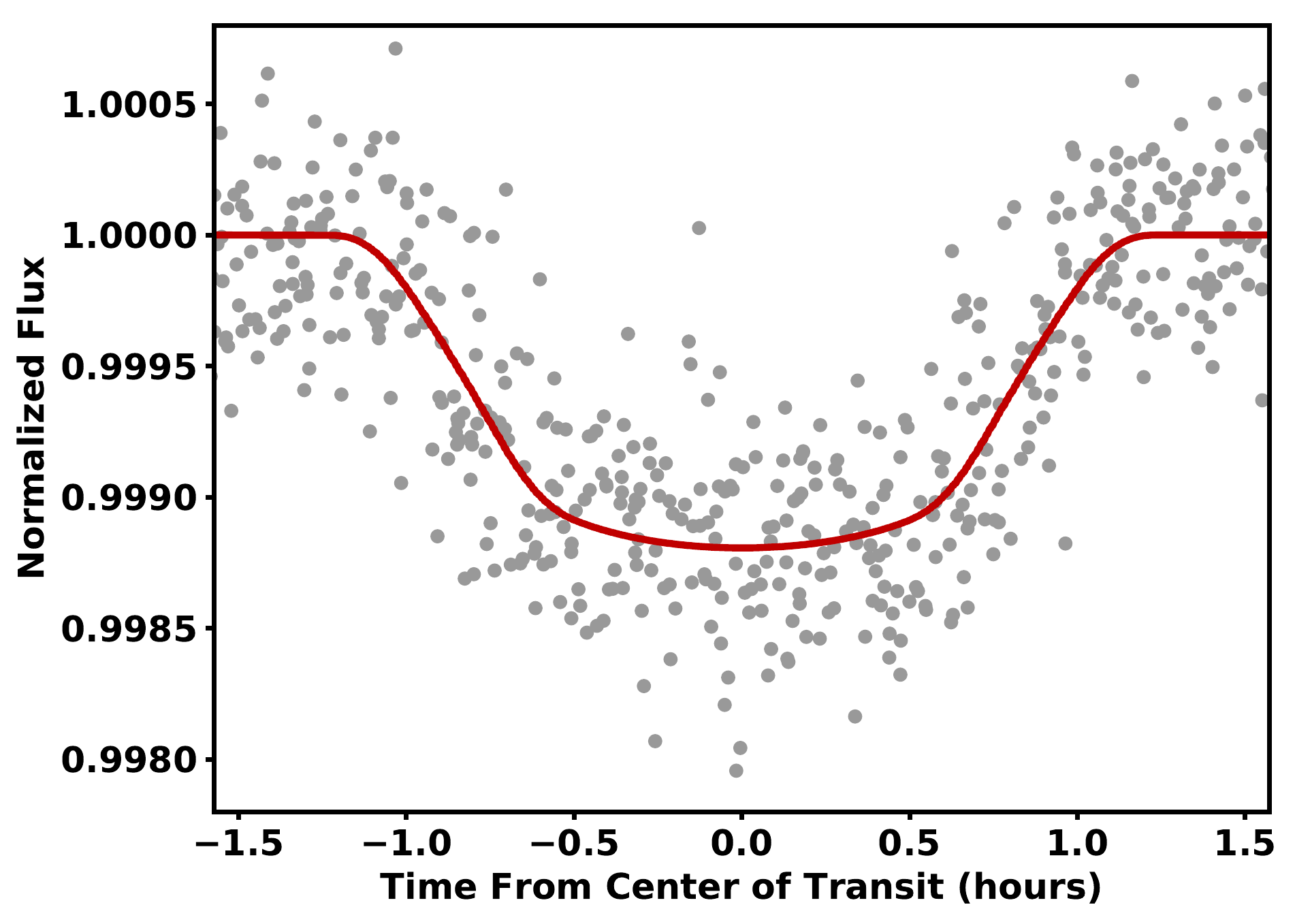}
    \caption{Phase-folded light curve of \starname\ (left) and \starnametwo\ (right) from \kepler\ (grey points) with the best-fit transit model (red). The best-fit GP model to the stellar variability has been removed from both the data and the model for clarity. Both planets show the expected transit shape.}
    \label{fig:ecctransit}
\end{figure*} 

For each of the four runs (two planets each with and without eccentricity), we ran the MCMC using 50 walkers for 250,000 steps including a burn-in of 20,000 steps. The total run was more than 50 times the autocorrelation time (for all fits), indicating that the steps were sufficient for convergence. 

The best-fit parameters with uncertainties for both fits can be found in Table \ref{tab:bestfitParams}. We also show the phase folded light curves in Figure~\ref{fig:ecctransit} for the preferred fit. As expected, the long baseline of data provided excellent constraints on $P$ and $T_0$, but the longer (30\,m) cadence yielded only weak constraints on impact parameter. This can be seen in the corner plot for the major transit-fit parameters (Figure \ref{fig:ecccorner}).

\begin{figure}[tbhp]
    \includegraphics[width=0.47\textwidth]{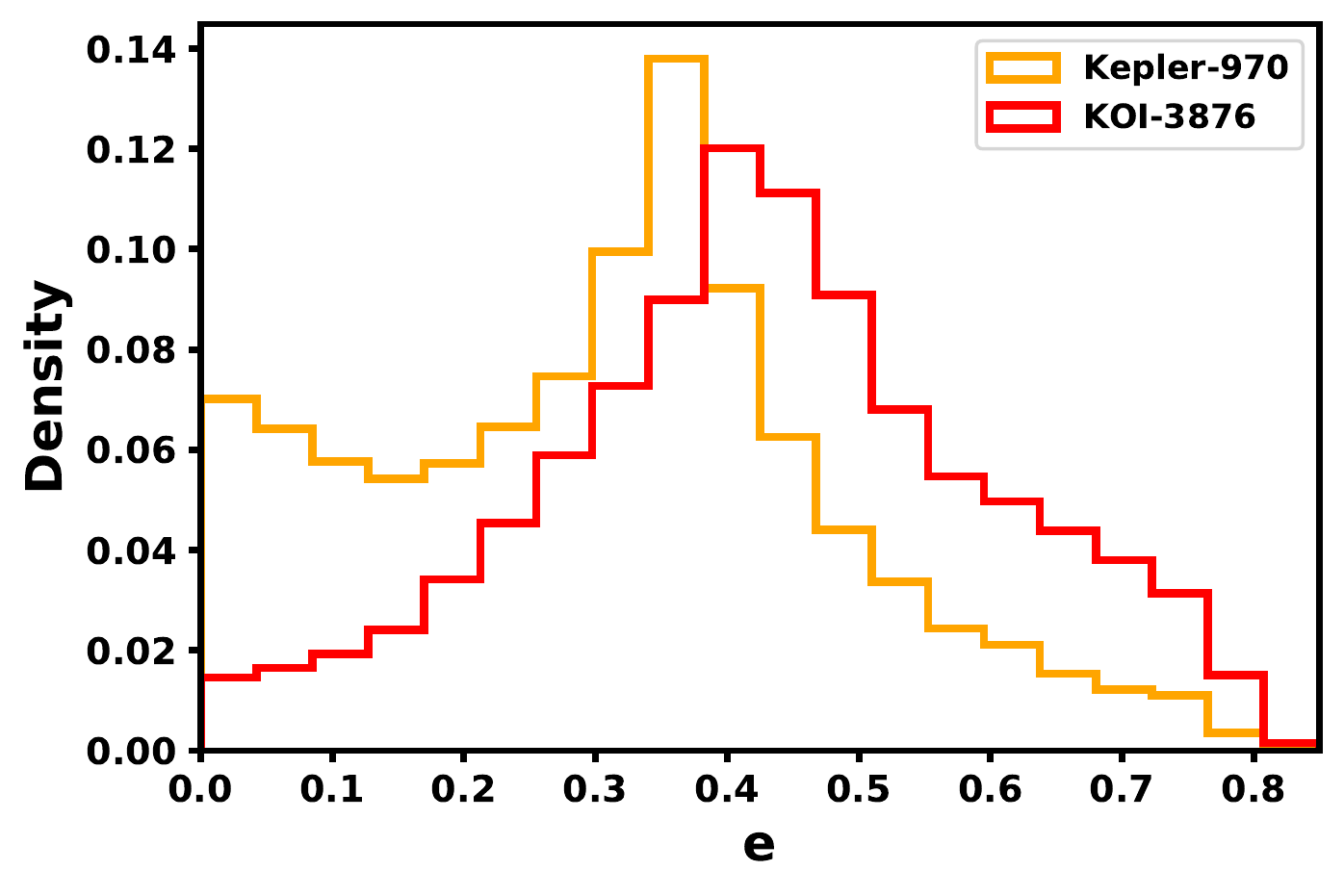}
    \caption{Posteriors of the eccentricity ($e$) for \planetname\ (red) and \planetnametwo\ (orange). For both systems, the $e=0$ fit yields a $\rho_\odot$ greater (shorter duration) than the expected value determined in Section~\ref{sec:stellar_params}. Uncertainties are large due to the long (30\,m) cadence, so the final eccentricity is consistent with zero in both cases (marginal consistency for \planetname). }
    \label{fig:ecchist}
\end{figure}

For \starname\ and \starnametwo, the first fit ($e=0$) yielded a $\rho_{\star}$ value larger than the spectroscopic/isochronal value determined in Section~\ref{sec:stellar_params} (15.5$\rho_\odot$ vs 1.3$\rho_\odot$ and 15.2$\rho_\odot$ vs 1.9$\rho_\odot$). Although the error on the transit-fit density is large (5.9$\rho_\odot$ and 6.7$\rho_\odot$), the two values were consistent at $\simeq2\sigma$. But it is suggestive that the planets are on eccentric orbits. Indeed, in the fits where $e$ is allowed to float, the highest-likelihood models had $e\gtrsim0.3$ for both planets, with the posteriors shown in Figure~\ref{fig:ecchist}. For this reason, we prefer the fit where $e$ was allowed to float for both targets.

\begin{figure*}[p]
    \centering
    \includegraphics[trim=50 12 50 35,clip=True,width=0.92\textwidth]{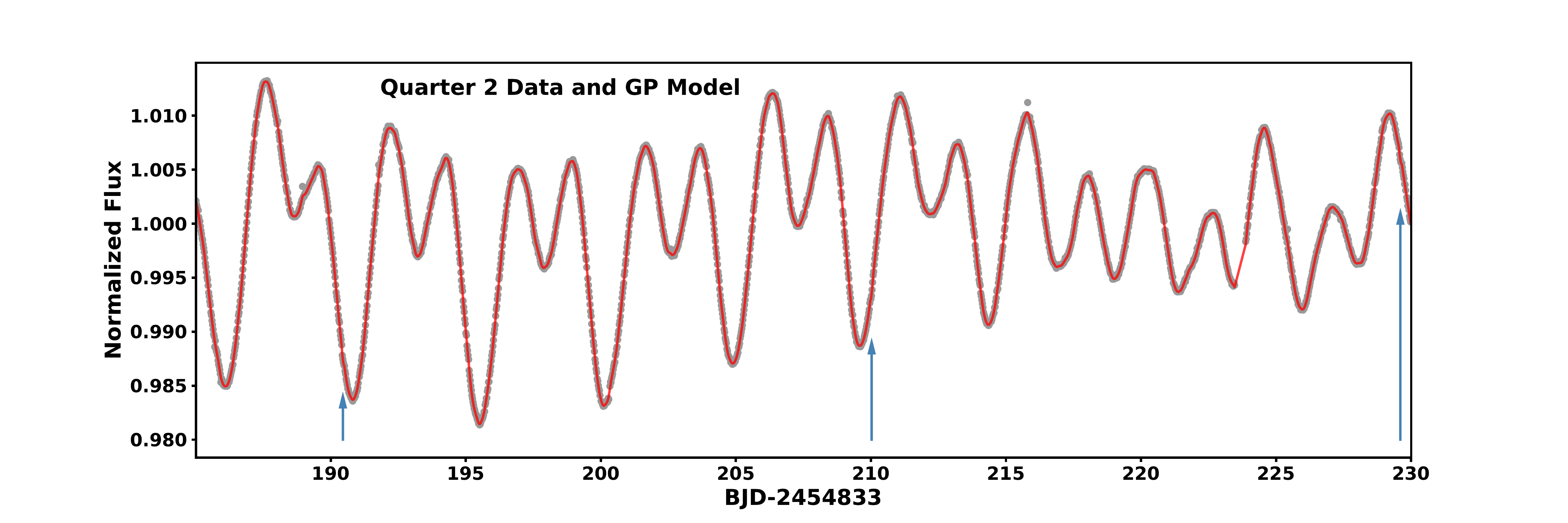}
    \includegraphics[trim=50 12 50 35,clip=True, width=0.92\textwidth]{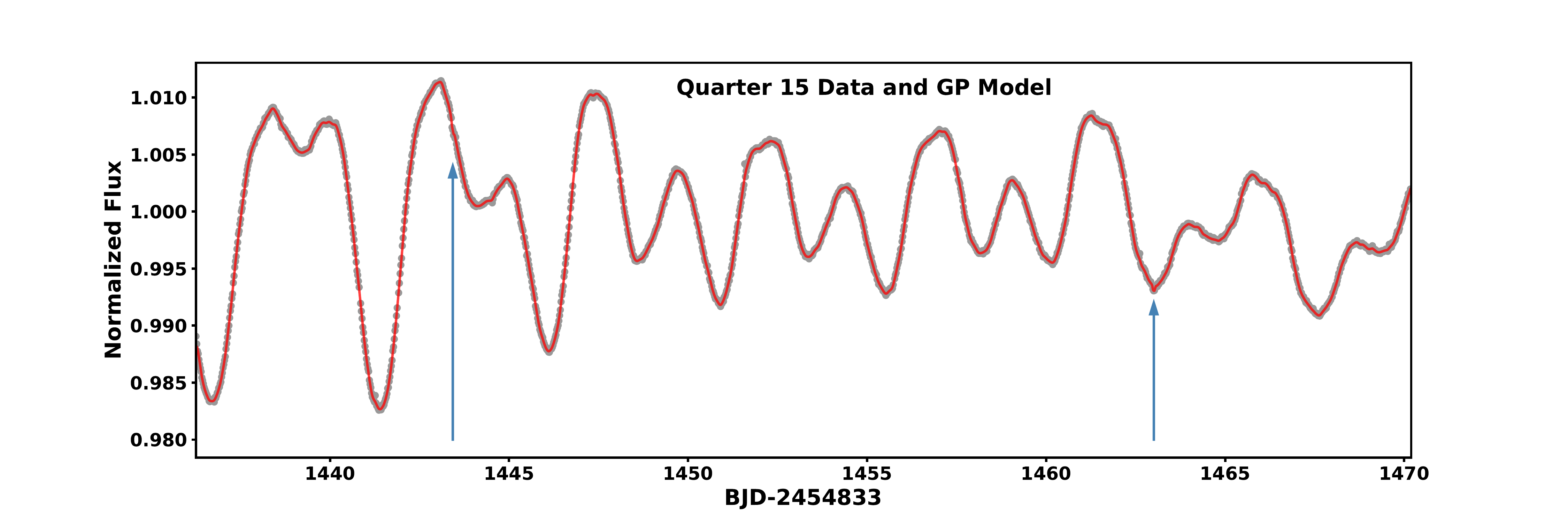}
    \includegraphics[trim=50 12 50 35,clip=True,width=0.92\textwidth]{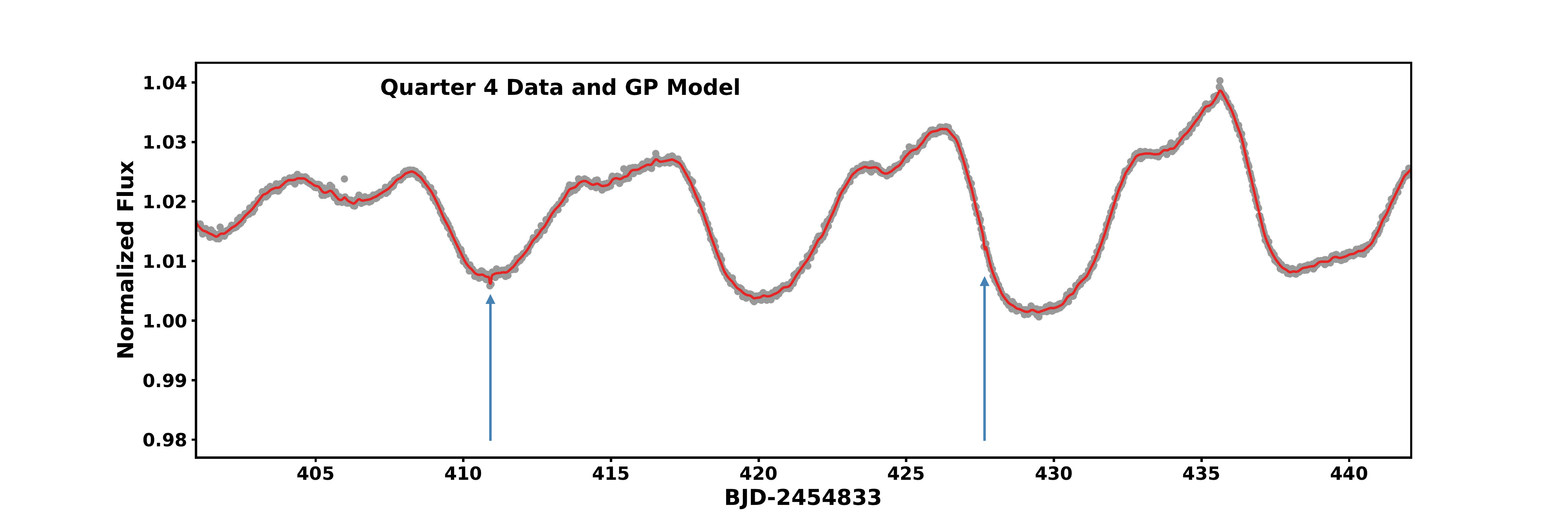}
    \includegraphics[trim=50 12 50 35,clip=True,width=0.92\textwidth]{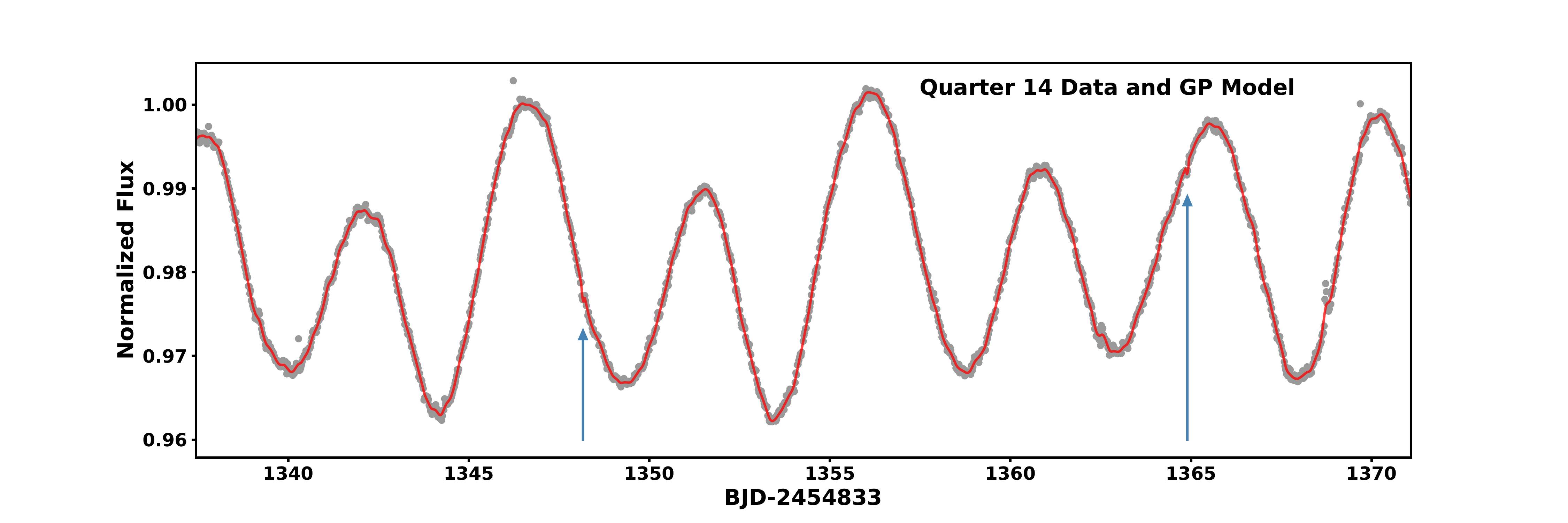}
    \caption{Representative sections of the two planet host light curves. The top two are for \starname, while the bottom two are for \starnametwo. The grey points are the \kepler\ data, the red line shows the best-fit GP model, and the arrows indicate transits. }
    \label{fig:quarter2GP}
\end{figure*}

The SHO GP fits did an excellent job describing the overall variability for both planets, even in the presence of complex changes in the light curve morphology over 4.5 years of observations by \kepler. We show two example quarters for each system in Figure~\ref{fig:quarter2GP} that highlight this.

\subsection{Transit timing variations in \planetnametwo}

\begin{figure}[tbh]
    \centering
    \includegraphics[width=0.45\textwidth]{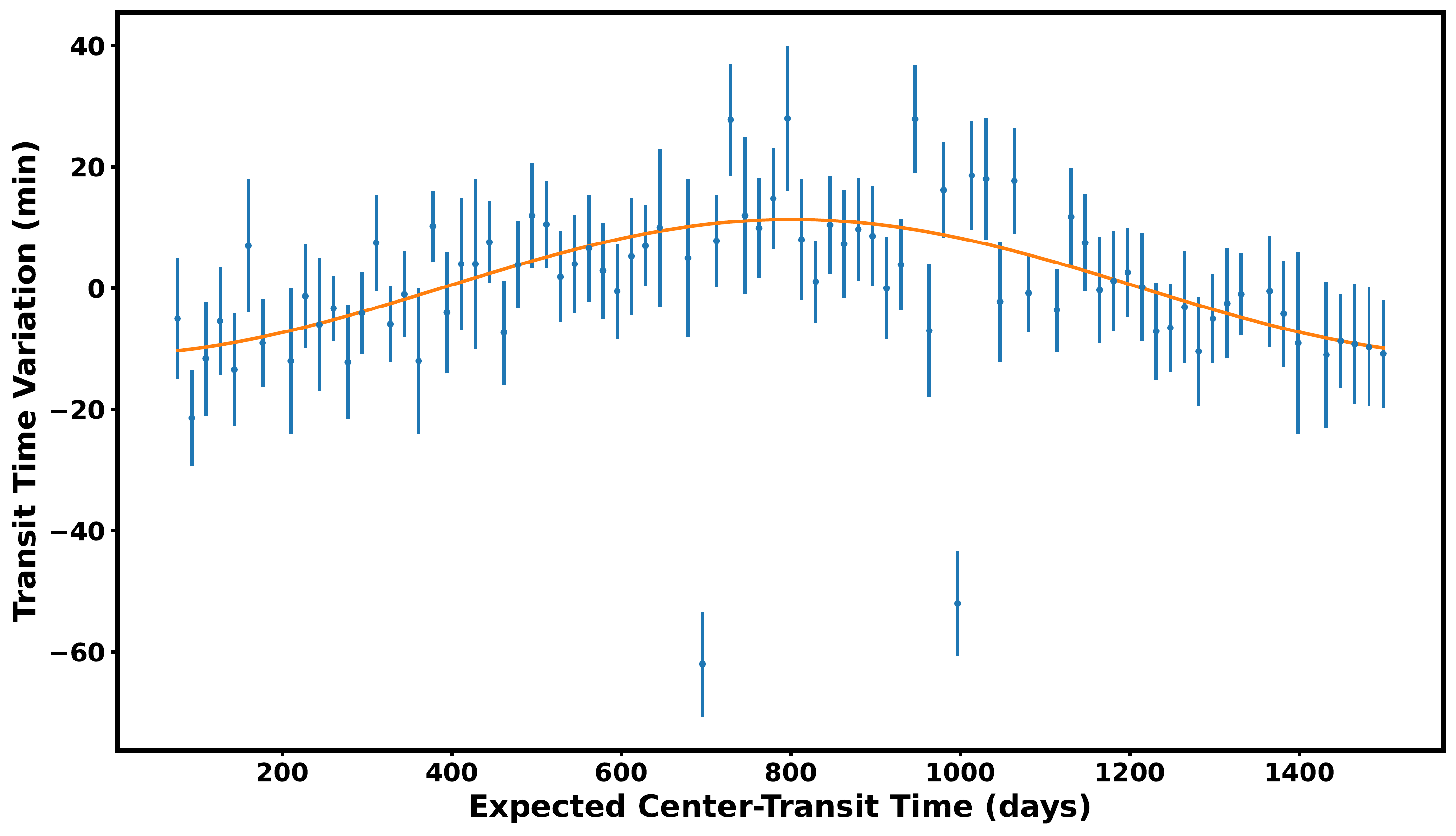}
    \caption{Transit timing of \planetnametwo\ from \citep{2016ApJS..225....9H}. The orange is our best-fit to the data assuming the variation is from a companion on a circular orbit. \label{fig:ttv}}
\end{figure}

\planetnametwo\ planet was previously shown to exhibit transit timing variations \citep[TTVs; e.g.,][]{2012ApJ...756..185F, 2016ApJS..225....9H}, which we show in Figure~\ref{fig:ttv}. The period implied by the variation is too long to be explained by a planet near resonance perturbing the orbit. As shown in \citet{Ioannidis2016}, it is rare for TTVs to be caused by starspots, especially given the signal here is unrelated to the stellar rotation signal. A more likely cause is a wider stellar or brown dwarf companion causing \starnametwo\ to change positions as it orbits the barycenter. 

To approximate the range of possible companions, we performed a least-squares fit to the TTVs from \citet{2016ApJS..225....9H}. We assumed a simple sine curve for a circular orbit, fitting the TTV amplitude and period, as well as two nuisance parameters capturing the phase and TTV offsets. We excluded two anomalous points from the fit (Figure~\ref{fig:ttv}). Unfortunately, the period is comparable to, or longer than the observing window of \kepler, so the resulting fit was highly uncertain, yielding a period of 1600$\pm$300\,days and a TTV semi-amplitude of 11$\pm$2\,minutes. Assuming an edge-on orbit, this corresponds to a companion with mass $0.06M_\odot<M_*<0.11M_\odot$. 

At 300\,pc, the corresponding companion would be too close to the parent star and faint to be detected by existing adaptive optics imaging of \starnametwo. Multi-year RVs could detect such a companion, but the target has only a single RV of sufficient precision (the CKS spectrum discussed in Section~\ref{sec:acceptedMems}). Fortunately, the possible companion would also be too faint to impact the transit properties or be responsible for the transit. We leave detection and further characterization of this candidate companion for a future effort.

\subsection{False Positive Analysis}\label{sec:fpp}
In \citet{2016ApJ...822...86M}, the authors run the false-positive probability calculator \texttt{VESPA} \citep{2015ascl.soft03011M} on all \kepler\ objects of interest available at the time, which included both \planetname\ and \planetnametwo. Kepler-970\,b was validated as planetary, and our qualitative inspection of the light curve and archival data re-affirmed this conclusion. The light curve has the expected shape, there are no visible companions in the high-resolution imaging, and the nearest star detected in \gaia\ imaging ($\simeq$4\arcsec\ away) is too faint to reproduce the transit depth and shape. We did not revisit this assessment. 

\begin{figure}[hbt]
    \centering
    \includegraphics[width=0.47\textwidth]{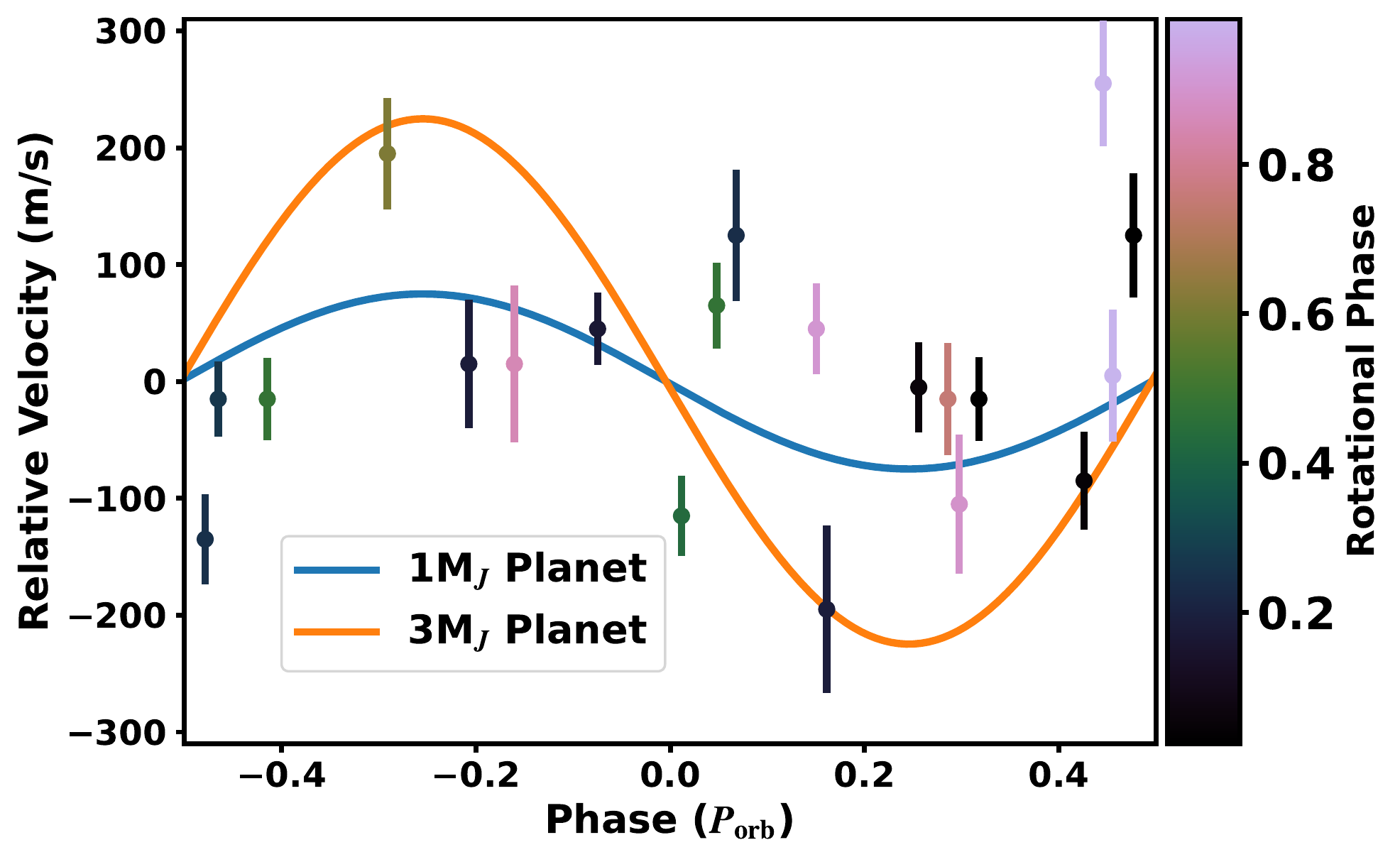}
    \caption{Radial velocities from APOGEE \citep{2020AJ....160..120J} for \starname\ as a function of the planet's orbital phase and colored by the rotational phase. The velocities rule out any companion more massive than $\simeq2M_J$, ruling out any possibility of an eclipsing binary at the transit period. The scatter is larger than expected for the uncertainties by $\simeq$100\mps, most likely due to stellar jitter common in young stars \citep{2019A&A...632A..37B, 2021AJ....161..173T}. The residual jitter is still far below the expected variation for a tight eclipsing binary. }
    \label{fig:RVs}
\end{figure} 

 However, \citet{2016ApJ...822...86M} assigned a high probability (90\%) that the signal associated with \planetname\ is an eclipsing binary and a $<1\%$ probability that the signal is due to a planet. This conclusion was based primarily on the light curve morphology and available stellar parameters. 

 As we show in Figure~\ref{fig:RVs}, radial velocities from the Apache Point Observatory Galactic Evolution Experiment 16th data release \citep[APOGEE DR16;][]{2020AJ....160..120J} rule out any stellar companion at the period of the planet. Further, our light curve analysis shows the expected U-shape transit for a planet, and there is no sign of a companion in the extant spectroscopy or adaptive optics imaging and non-redundant aperture masking from \citet{Kraus2016a}. \gaia\ EDR3 astrometry and imaging similarly show no sign of a companion. There is only one star detected with the \kepler\ PSF, which is too faint to reproduce the transit. \starname\ has a low Renormalised Unit Weight Error (RUWE) in EDR3 (0.94). RUWE value is effectively an astrometric reduced $\chi^2$ value, normalized to correct for color and brightness dependent effects\footnote{\url{https://gea.esac.esa.int/archive/documentation/GDR2/Gaia_archive/chap_datamodel/sec_dm_main_tables/ssec_dm_ruwe.html}}. RUWE should be around 1 for well-behaved sources, and higher values (RUWE$\gtrsim$1.3) suggests with the presence of a stellar companion \citep{Ziegler2020, Wood2021}. 
 
 It is possible the high false-positive probability from \citet{2016ApJ...822...86M} was an artifact of poor detrending of the high stellar variability in \starname\ and/or the mismatch between the transit duration and that expected for a circular orbit (see Section~\ref{sec:transit}). We re-ran the \texttt{VESPA} analysis using our GP-detrended curve and the updated imaging constraints. We found a 97\% probability \planetname\ is a planet and a 3\% probability \planetname\ is an eclipsing binary. Other false-positive scenarios (background EB and hierarchical EB) had negligible ($<0.1$\%) probabilities. As previously stated and shown in Figure \ref{fig:RVs}, the APOGEE velocities rule out a stellar companion to \starname\ at the period of the planet. Including the radial velocities eliminates the EB scenario and reduces the FPP to below 0.1\%, validating the signal as planetary in nature.

\section{Summary and Conclusions}\label{sec:summary}

\subsection{\association}
\association\ is a young ($105\pm10$\,Myr) association that overlaps with the \kepler\ field. We initially identified the association as an overdensity of stars in Galactic positions and tangential velocities around \starname. Through rotation periods and lithium abundances of candidate members, and a comparison to model isochrones, we showed the collection of stars is coeval, with an age similar to that of Pleiades ($\simeq$ 110 Myr). We refered to this group as \association, although it is likely one part of the known Theia 316 string \citep{2019AJ....158..122K, 2020AJ....160..279K}. 

Interestingly, even among our high-probability members of \association, more than half were missing from the Theia 316 list. Similarly, our list only contained one branch of Theia 316. The structure likely harbors far more stars than even the combination of the two lists. A more detailed analysis with \tess\ rotations and \gaia\ DR3 velocities would be invaluable here, and similar analysis the thousands of other candidate associations identified by \citet{2020AJ....160..279K}.

Stellar angular momentum and rotationally-driven magnetic activity are both age and mass-dependent. Rotation periods in open clusters are useful for studying these phenomena, but there is a lack of low-mass stars ($<$0.5 M$_\odot$) with rotation periods available \citep{Covey2016}. Of the rotation periods in \association\ utilized here, 49 are K- or M-dwarfs stars, which are more likely to be within this low-mass boundary. Further work on the analysis of the distribution of rotation periods versus color can be done and extend the sample size used in analyses, such as in \citet{2016AJ....152..114R}. 

The long-baseline of \kepler\ data has enabled studies of spot lifetimes \citep[e.g.,][]{2017MNRAS.472.1618G}. For example, starspots have shorter lifetimes on stars with faster rotation periods \citep{2022ApJ...924...31B}. This is likely due to age, i.e., that younger stars have shorter spot lifetimes.  \association\ and the younger $\delta$-Lyr cluster \citep{2021arXiv211214776B} have enough overlap with the \kepler\ field for a more direct test of how spot lifetimes (among other properties) vary with age. 

\subsection{\planetname\ and \planetnametwo}
Taking into account the updated (younger) age and utilizing new methods for analyzing variable light curves, we updated the stellar and transit parameters for \planetname. We found \planetname\ to be about twice the size of the earth, orbiting a young analog to the Sun every 19.577 days. The transit duration also suggests a modest eccentricity, although errors are too large to rule out a circular orbit. \citet{2016ApJ...822...86M} previously found \planetname\ likely to be an eclipsing binary, we rule out this disposition based on APOGEE radial velocities and our light curve analysis and validate the signal as planetary in origin.

\planetnametwo\ was previously validated as a planet \citep{2016ApJ...822...86M}. As with \planetname, we update the stellar and planet parameters using the new younger age. We find \planetnametwo\ to be about 2.5$R_\oplus$, orbiting a young K dwarf every 16.74 days. 

We searched for additional transiting planet candidates in \association. Although we identify many candidates and recover some known KOIs (see Table~\ref{tab:kois-kics}), most of these were either not real members, false positives, or did not meet our threshold for significance. Ultimately, only \planetname\ and \planetnametwo\ passed all tests. Additional \tess\ data of the \kepler\ field, as well as a more complete census of the Theia 316 group, might yield more planets.

\begin{figure}[tbh]
    \centering
    \includegraphics[width=0.47\textwidth]{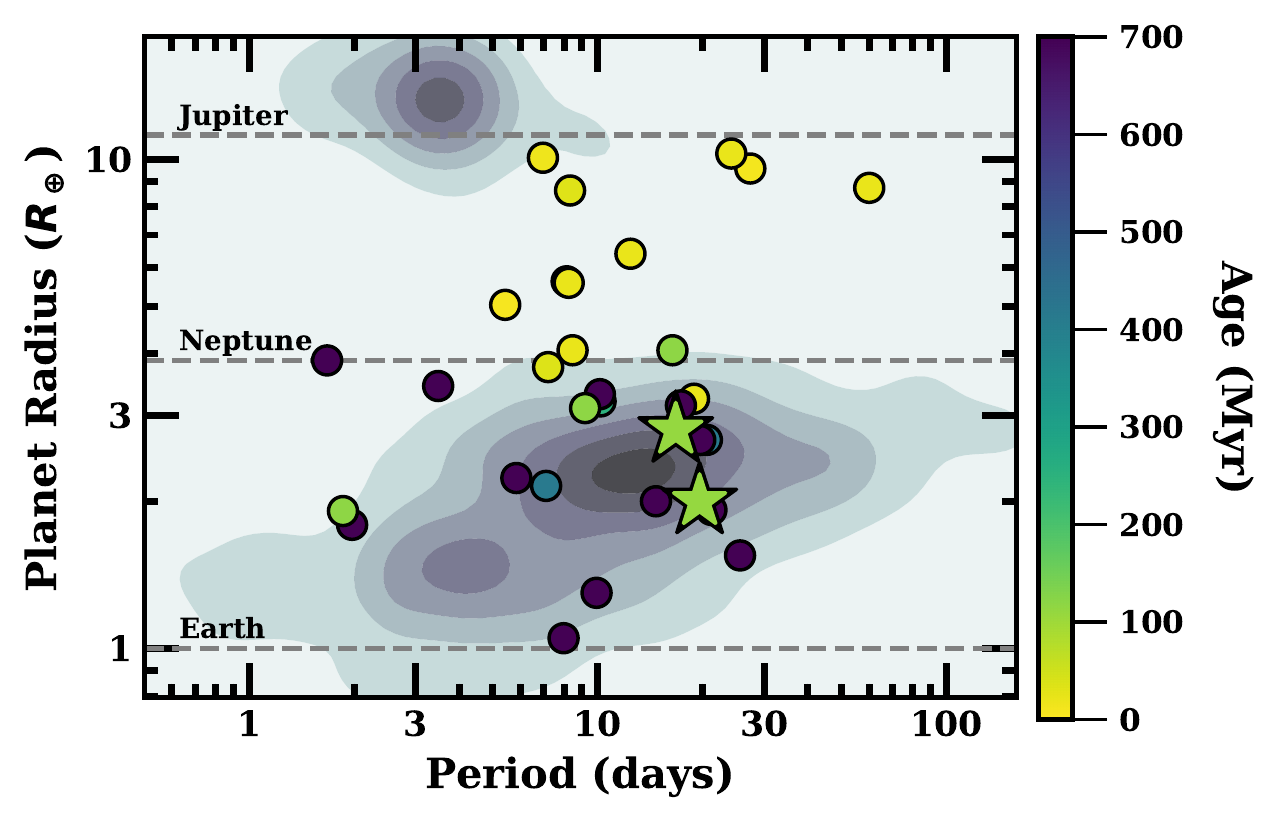}
    \caption{The current census of young (Hyades-age or younger) transiting planets within clusters or young associations. Points are color-coded by age. The two stars are \planetname\ and \planetnametwo. The contours represent the density of (mostly older) transiting planets from \kepler. Planet parameters taken from the NASA exoplanet archive \citep{NASAexplanetarchive}.}  
    \label{fig:census}
\end{figure}

These two planets join the growing number of young transiting planets, which we show in Figure~\ref{fig:census}. \planetname\ and \planetnametwo\ land in the heavily populated region of mini-Neptunes. Both have similar or smaller radii and similar or longer periods than the similar-aged objects found by \ktwo\ and \tess. They also sit well below the infant (10-50\,Myr) planets, which tend to land between Neptune and Jupiter due to a combination of radius evolution \citep{Mann2017a, 2020MNRAS.498.5030O} and sensitivity variations with age \citep{Rizzuto2017}.

Recent searches found no transiting planets in the \ktwo\ data of the similar-aged Pleiades cluster \citep{Gaidos:2017aa, Rizzuto2017}. \citet{Gaidos:2017aa} argues this deficiency is not surprising given our sensitivity to planets in the Pleiades, although the more effective transit-search methods of \citet{Rizzuto2017} found mild tension with the planet occurrence seen in Pleiades compared to the older Hyades and Praesepe. The discovery of TOI-451\,bcd in the $\simeq$120\,Myr Psc-Eri cluster \citep{THYMEIV} and the planets in this work suggest the lack of planets in the Pleiades was not related to age. We inspected the sensitivity results from \citet{Rizzuto2017} and concluded that the planets presented in this work would have landed just below the detection limits of the \ktwo\ data of similar stars in the Pleiades. 

We have known there are young planet-hosts in the \kepler\ field from their rotation periods \citep{2013MNRAS.436.1883W, 2021AJ....161..265D} and lithium levels \citep{2018ApJ...855..115B}, but such ages are generally imprecise when applied to individual stars. The previous challenge for \kepler\ was the lack of young associations. \association\ and the recently identified $\delta$-Lyr cluster \citep{2021arXiv211214776B} demonstrate that the previous list of just four clusters in the \kepler\ field was incomplete and motivates the work for further searches in the \kepler, \ktwo, and CoRoT \citep{2009A&A...506..411A} fields. In addition to surveys like Theia \citep{2019AJ....158..122K, 2020AJ....160..279K}, the methods applied here may reveal a new population of young associations harboring known transiting systems. 

\begin{acknowledgements}
The authors thank the anonymous referee for their feedback on the paper. The authors also thank Erik Petigura, Luke Bouma, and Tim Bedding for their useful discussions on this manuscript. We also would like to acknowledge Halee, who kept Madyson sane during the lockdown. 

During the preparation of this manuscript, KOI-3876 was independently validated by \citet{Valizadegan2022} and assigned the name Kepler-1928. 

MGB was supported by the NC Space Grant Undergraduate Research program and by funding from the Chancellor's Science Scholars Program at the University of North Carolina at Chapel Hill. AWM was supported through a grant from NASA’s Exoplanet Research Program (XRP; 80NSSC21K0393). This material is based upon work supported by the National Science Foundation Graduate Research Fellowship Program under Grant No. DGE-1650116 to PCT.

This paper includes data collected by the TESS mission, which are publicly available from the Mikulski Archive for Space Telescopes (MAST). Funding for the TESS mission is provided by NASA’s Science Mission directorate. This research has made use of the Exoplanet Follow-up Observation Program website, which is operated by the California Institute of Technology, under contract with the National Aeronautics and Space Administration under the Exoplanet Exploration Program. This work has made use of data from the European Space Agency (ESA) mission \emph{Gaia} \footnote{\url{https://www.cosmos.esa.int/gaia}}, processed by the \emph{Gaia} Data Processing and Analysis Consortium (DPAC)\footnote{\url{https://www.cosmos.esa.int/web/gaia/dpac/consortium}}. Funding for the DPAC has been provided by national institutions, in particular, the institutions participating in the \emph{Gaia} Multilateral Agreement.  This research has made use of the VizieR catalogue access tool, CDS, Strasbourg, France. The original description of the VizieR service was published in A\&AS 143, 23. Resources supporting this work were provided by the NASA High-End Computing (HEC) Program through the NASA Advanced Supercomputing (NAS) Division at Ames Research Center for the production of the SPOC data products. 

\end{acknowledgements}

\vspace{5mm}
\facilities{TESS, Kepler, Gaia, Sloan (APOGEE), Smith (Coude)}

\software{\texttt{misttborn.py} \citep{MISTTBORN}, \texttt{galpy} \citep{2015ApJS..216...29B}, \textit{emcee} \citep{Foreman-Mackey2013}, \textit{batman} \citep{Kreidberg2015}, matplotlib \citep{hunter2007matplotlib}, \texttt{corner.py} \citep{foreman2016corner}, \texttt{Celerite} \citep{celerite}
}

\bibliography{fullbiblio}{}
\bibliographystyle{aasjournal}

\clearpage

\begin{longrotatetable}
\begin{deluxetable*}{cccccccccccccccccc}\label{tab:fullsample}
\tablehead{
\colhead{Gaia EDR3} & \colhead{$\alpha$} & \colhead{$\delta$} & \colhead{Gmag} & \colhead{$V_{\rm{off}}$} & \colhead{Spectral\tablenotemark{a}} & \colhead{TESS\tablenotemark{b}} & \colhead{Kepler\tablenotemark{b}} & \colhead{$\pi$} & \colhead{$\sigma_\pi$} & \colhead{$P_{\rm{rot}}$} & \colhead{$\sigma_{P_{\rm{rot}}}$} & \colhead{$P_{\rm{rot}}$\tablenotemark{c}} & \colhead{Li} & \colhead{$\sigma_{\rm{li}}$} & \colhead{RV} & \colhead{$\sigma_{\rm{RV}}$} & \colhead{RV\tablenotemark{d}}\\
\colhead{} & \colhead{(J2016)} & \colhead{(J2016)} & \colhead{(mag)} & \colhead{(km/s)} & \colhead{Class} & \colhead{} & \colhead{} & \colhead{(mas)} & \colhead{(mas)} & \colhead{(days)} & \colhead{(days)} & \colhead{ref} & \colhead{(m\AA)} & \colhead{(m\AA)} & \colhead{(km/s)} & \colhead{(km/s)} & \colhead{ref}
}
\startdata
2052827207364859264 & 290.44063 & 38.52357 & 12.605 & 0.000 & G7.3 & Y & Y & 3.057 & 0.009 & 4.665 & 0.035 & 2, 3 & 134.0 & 18.0 & -26.09 & 0.03 & 5 \\
2101379205604338688 & 289.62522 & 40.70874 & 14.788 & 3.102 & K5.3 & N & Y & 3.015 & 0.019 & 9.23 & 0.66 & 3 & $<$40 & $\ldots$ & -27.15 & 0.1 & 9 \\
2052804323776522624 & 290.46921 & 38.20201 & 18.977 & 3.554 & M3.9 & N & N & 3.036 & 0.172 & $\ldots$ & $\ldots$ & $\ldots$ & $\ldots$ & $\ldots$ & $\ldots$ & $\ldots$ & $\ldots$ \\
2100939194794324608 & 289.80783 & 38.95042 & 10.829 & 0.478 & F4.4 & Y & N & 3.024 & 0.012 & $\ldots$ & $\ldots$ & $\ldots$ & $<$20 & $\ldots$ & -27.03 & 0.18 & 5 \\
2052954995531623040 & 291.28017 & 39.20070 & 14.355 & 4.348 & K4.7 & N & Y & 3.055 & 0.016 & 13.849 & 0.035 & 1, 2, 3 & $\ldots$ & $\ldots$ & $\ldots$ & $\ldots$ & $\ldots$ \\
2100967537279486336 & 290.14895 & 39.26079 & 18.451 & 3.806 & M4.1 & N & N & 3.094 & 0.136 & $\ldots$ & $\ldots$ & $\ldots$ & $\ldots$ & $\ldots$ & $\ldots$ & $\ldots$ & $\ldots$ \\
2051102868195719168 & 290.43983 & 37.73210 & 17.503 & 0.516 & M3.2 & N & N & 3.019 & 0.069 & $\ldots$ & $\ldots$ & $\ldots$ & $\ldots$ & $\ldots$ & $\ldots$ & $\ldots$ & $\ldots$ \\
2053037660761796096 & 290.86832 & 39.65124 & 17.902 & 0.605 & M3.7 & N & N & 3.067 & 0.084 & $\ldots$ & $\ldots$ & $\ldots$ & $\ldots$ & $\ldots$ & $\ldots$ & $\ldots$ & $\ldots$ \\
\enddata
\tablecomments{Table is truncated. The full version will be available online.}
\tablenotetext{a}{Spectral types are assigned based on the star's $B_P-R_P$ color and the table from \citet{2013ApJS..208....9P}.
}

\tablenotetext{b}{Indicates if the target had a light curve in \tess\ or \kepler\ of sufficient quality for running the Notch transit-search pipeline. }
\tablenotetext{c}{Key for rotation period references: 1 = \citep{McQuillan2013, 2014ApJS..211...24M}, 2 = \citep{Nielsen:2013}, 3 = \citet{Santos2021,Santos2019}, 4 = This work}
\tablenotetext{d}{Key for radial velocity references: 5 = Coude, 6 = LAMOSTDR5 \citep{2019yCat.5164....0L}, 7 = Gaia DR2 \citep{DR2_velocities}, 8 = APOGEE DR16 \citep{2020AJ....160..120J}, 9 = CKS \citep{CKS_cool}}
\end{deluxetable*}
\end{longrotatetable}

\end{document}